\documentclass[aps,twocolumn,amsmath,amssymb]{revtex4}

\usepackage{graphicx}
\usepackage{dcolumn}
\usepackage{bm}
\usepackage{xcolor}

\begin{document}


\title{The fate of time-reversal symmetry breaking in UTe$_2$}

\author{M.O. Ajeesh}
\author{M. Bordelon}
\author{C. Girod}
\author{S. Mishra}
\author{F. Ronning}
\author{E.D. Bauer}
\author{B. Maiorov}
\author{J.D. Thompson}
\author{P.F.S. Rosa}
\author{S.M. Thomas}
\email{smthomas@lanl.gov}
\affiliation{Los Alamos National Laboratory, Los Alamos, NM 87545}

\date{\today}

\begin{abstract}
Topological superconductivity is a long-sought state of matter in bulk materials, and odd-parity superconductor UTe$_2$ is a prime candidate.
The recent observation of a field-trainable spontaneous Kerr signal in UTe$_2$ at the onset of superconductivity provides strong evidence that the superconducting order parameter is multicomponent and breaks time-reversal symmetry.
Here, we perform Kerr effect measurements on a number of UTe$_2$ samples --- grown $via$ both chemical vapor transport and the molten-salt-flux methods --- that show a single superconducting transition between 1.6~K and 2.1~K.
Our results show no evidence for a spontaneous Kerr signal in zero field measurements.
This implies that the superconducting state of UTe$_2$ does not intrinsically break time-reversal symmetry.
Instead, we observe a field-trainable signal that varies in magnitude between samples and between different locations on a single sample, which is a sign of inhomogeneous magnetic regions.
Our results provide an examination of representative UTe$_2$ samples and place strong constraints on the superconducting order parameter of UTe$_2$.
\end{abstract}

\maketitle

\subsection{Introduction}

In a superconductor, electrons form pairs that condense into a macroscopic quantum state with zero electrical resistance.
The underlying symmetries of the pair wavefunction are important to determine the properties of the superconducting state.
When the orbital wavefunction is antisymmetric [$\phi(k) = -\phi(-k)$], odd-parity superconductivity is realized~\cite{Sigrist1991}.
When time-reversal symmetry is broken, chiral superconductivity may emerge. Odd-parity chiral superconductors are sought-after materials predicted to host topological excitations with nonabelian statistics that could enable quantum computing~\cite{Kallin2016a,Sato2017a}.
Over the past decade, the search for an odd-parity chiral superconductor has intensified, but an ideal candidate has yet to be established.
Sr$_2$RuO$_4$ was long thought to be the leading candidate~\cite{Mackenzie2017}, but recent nuclear magnetic resonance experiments showed that the superconducting order parameter is even parity~\cite{Pustogow2019}.

Odd-parity UTe$_2$ is a recently-discovered chiral superconductor candidate~\cite{Ran2019}.
Evidence for chiral superconductivity in UTe$_2$ comes from multiple experiments.
Scanning tunneling microscopy reveals an asymmetry in the in-gap surface states at step edges argued to arise from a chiral superconducting state~\cite{Jiao2019}.
Microwave surface impedance~\cite{Bae2021}  and anisotropic penetration depth~\cite{Ishihara2021} studies also support a chiral superconducting state.
In addition, a subset of specific heat measurements found a double peak near the superconducting transition temperature (T$_c$), which is similar to findings in chiral superconductor candidate UPt$_3$~\cite{Joynt2002,Schemm2014,Avers2020} and suggests the presence of a multicomponent superconducting state.
A key characteristic of a chiral superconductor is that its order parameter breaks time-reversal symmetry, and the polar Kerr effect (PKE) is a powerful probe of time-reversal symmetry breaking.
Previous PKE measurements in UTe$_2$ indeed found a spontaneous Kerr rotation that arises at the onset of superconductivity~\cite{Hayes2020}.
Because the Kerr rotation was found to be trainable by cooling in a $c$ axis magnetic field ($B_{1g}$ symmetry), a $B_{3u}+iB_{2u}$ superconducting order parameter was put forward as the most likely multicomponent state~\cite{Hayes2020}.
Further measurements of the PKE under larger training fields revealed that the magnitude of the Kerr rotation ($\theta_K$) scales with the training field up to a certain field~\cite{Wei2022}, which was attributed to a critical state of ferromagnetic vortices.

\begin{table*}[t]
	\caption{\label{tab:properties}Properties of samples studied via PKE including: sample name, growth method, superconducting transition temperature, the size of the specific heat jump at T$_c$, the residual specific heat determined from a linear fit of C/T versus T$^2$, and the residual resistivity ratio (see text).
	}
	\begin{ruledtabular}
	\begin{tabular}{lllllll}
		Sample & Growth Method & T$_c$ (K) & $\Delta{}$C/T$_c$ (mJ mol$^{-1}$ K$^{-2}$) & $\gamma{}^*$ (mJ mol$^{-1}$ K$^{-2}$) & RRR\\
		S1 & MSF & 2.1 &  266 & 6 & 400 \\
		S2 & CVT & 2.0 &  232 & 21 & 111 \\
		S3 & CVT & 1.9 &  163 & 41 & 79 \\
		S4 & CVT & 1.6 &  154 & 61 & 46 \\
	\end{tabular}
	\end{ruledtabular}
\end{table*}

Importantly, the properties of UTe$_2$ are markedly sensitive to sample growth conditions.
For samples grown $via$ chemical vapor transport (CVT), lower growth temperatures increase T$_c$ from 1.6~K to 2~K~\cite{Rosa2022}.
Further decreasing the growth temperature leads to a sudden disappearance of superconductivity and a drastic decrease in the residual resistivity ratio (RRR).
CVT growth either at higher temperatures or near the low-temperature limit often leads to samples with an apparent double transition in specific heat~\cite{Weiland2022}.
The determination of the origin of the double transition has been hampered by strong sample dependence of the superconducting state of UTe$_2$.
However, several studies now suggest that, unlike UPt$_3$, the double transition is not an intrinsic feature but rather a consequence of sample inhomogeneity~\cite{Thomas2021,Aoki2022}.
Further, single-transition samples show no evidence for a splitting of the superconducting transition under $B_{1g}$ shear stress that couples to the proposed $B_{3u}+iB_{2u}$ order parameter~\cite{Girod2022}, suggesting that the superconducting order parameter may be single component or belong to different symmetry channels.
More recently, the quality of UTe$_2$ single crystals was further improved using a molten-salt-flux (MSF) growth technique~\cite{Sakai2022}, leading to RRRs as large as 1000 and enabling the first observation of de Haas-van Alphen oscillations in UTe$_2$~\cite{Aoki2022b}.

A key outstanding open question is whether a spontaneous Kerr rotation persists in samples that show a single T$_c$.
PKE experiments to date have only been reported on a sample with a double transition near 1.55~K in specific heat~\cite{Hayes2020,Wei2022}.
In these samples, a large residual specific heat ($\gamma^*$) was observed in the superconducting state~\cite{Ran2019}, which has been shown to decrease as T$_c$ is increased~\cite{Cairns2020,Rosa2022,Aoki2022}.
Notably, a correlation has been identified between the volume fraction of inhomogeneous magnetic clusters detected by muon spin resonance ($\mu$SR) and the size of $\gamma^*$~\cite{Sundar2022}, which naturally points to the role of magnetism in samples with large $\gamma^*$.

Here, we investigate the polar Kerr effect on a number of samples grown via both the CVT and molten-salt methods.
All samples have only a single detectable transition in specific heat.
Our measurements do not show evidence for a spontaneous PKE effect that emerges at T$_c$ in zero-field-cooled measurements. 
Nonetheless, we observe a field-trainable signal that persists up to T$_c$ and whose magnitude varies significantly between different samples and between different spots on the same sample.
Contrary to the correlation observed in $\mu$SR~\cite{Sundar2022}, no clear trend is observed between the magnitude of the PKE and $\gamma^*$, which suggests that magnetic clusters may not be the only source of PKE.
Ac susceptibility measurements on another representative set of samples indicate a correlation between T$_c$ and the size of the vortex peak effect, a measure of crystalline quality.
This result unambiguously shows that lower-T$_c$ samples have a higher density of vortex pinning centers. 
Our results demonstrate that UTe$_2$ does not have a time-reversal symmetry breaking (TRSB) superconducting order parameter and that the inhomogeneous field-trainable PKE has an extrinsic origin.

\subsection{Results}

\begin{figure}[]
	\includegraphics[width=\columnwidth]{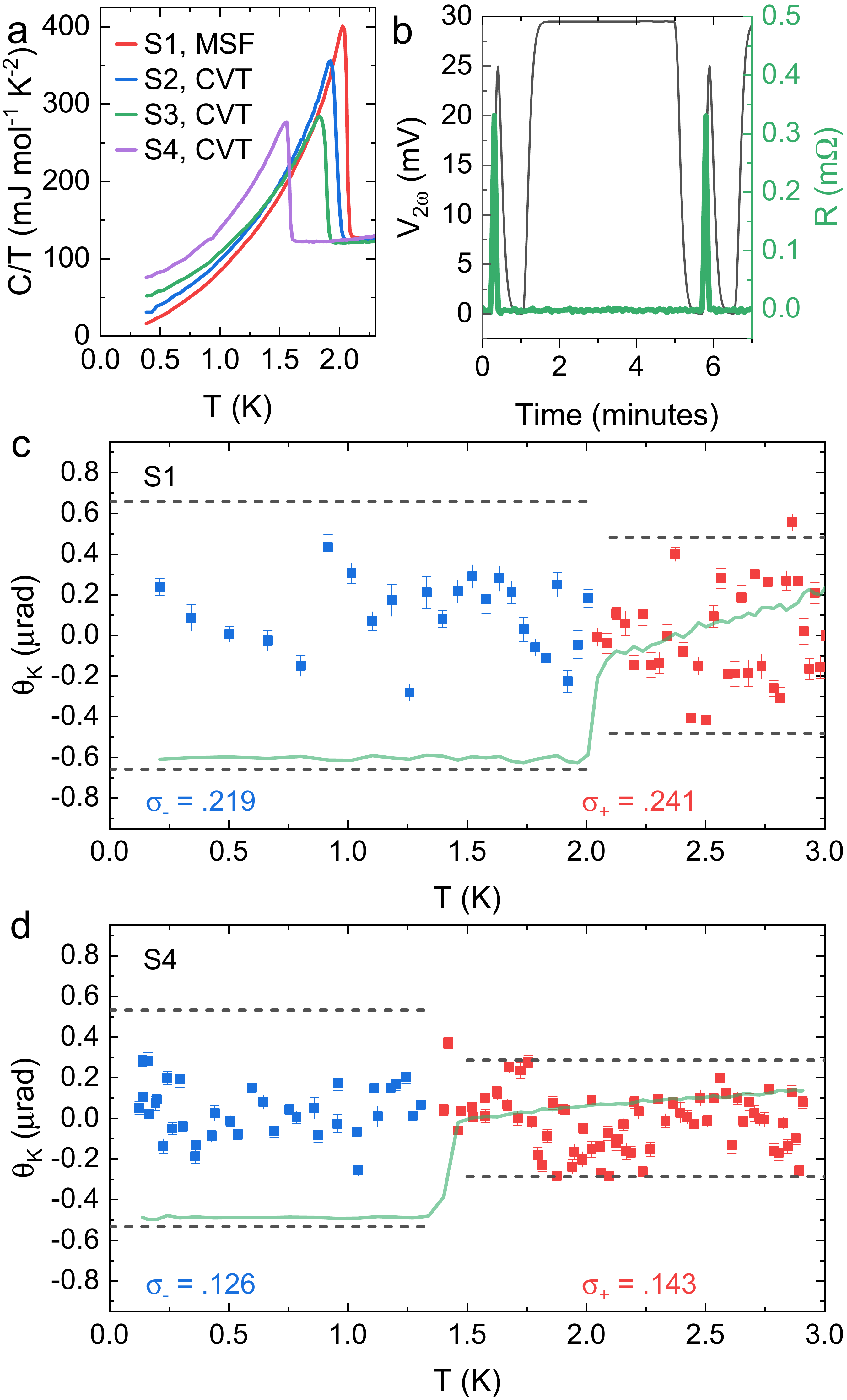}
    \caption{
		(a)~Specific heat of samples S1--S4.
		(b)~Example of laser pulse sequence used to heat the sample above T$_c$. The left axis shows V$_{2\omega}$ (proportional to light received by the detector), and the right axis shows the sample resistance.
		(c),(d)~Measured value of $\theta{}_K$ as a function of temperature for sample S1 and S4, respectively. The sample is heated above T$_c$ to allow for the possibility of different domain configurations between each data point. Red indicates T$>$T$_c$; blue indicates T$<$T$_c$. The faint green line shows the sample resistivity. The dashed lines indicate the expected increase in $\sigma$ below T$_c$ based on the observation in Wei \textit{et al.} (see text)~\cite{Wei2022}.
		}
    \label{fig1}
\end{figure}

Temperature-dependent specific heat measurements for each of the samples on which PKE was measured, shown in Fig.~\ref{fig1}a,
reveal a single superconducting transition with a transition width of approximately 50~mK.
T$_c$ varies from a minimum of 1.6~K in sample S4 to a maximum of 2.1~K in sample S1.
Samples S2--S4 were grown via chemical vapor transport~\cite{Rosa2022}, whereas sample S1 was grown using the molten-salt-flux method~\cite{Sakai2022}.
As observed previously, the $\gamma^*$ decreases dramatically as T$_c$ increases~\cite{Rosa2022,Cairns2020,Aoki2022}, and the MSF sample exhibits the smallest $\gamma^*$ of only 6 mJ~mol$^{-1}$~K$^{-2}$~\cite{Sakai2022}.
Table~I summarizes key properties of the samples investigated here.

Figures~\ref{fig1}c-d show the temperature-dependent polar Kerr rotation of our zero-field cooling experiments for samples S1 and S4, respectively.
To ensure zero-field cooling conditions, each sample was heated above $T_c$ using a high-intensity light pulse as shown in Fig.~\ref{fig1}b (see Methods for details).
Each data point in Figures~\ref{fig1}c-d represents the mean of $\theta_K$ measured after the sample is cooled to the indicated temperature from higher temperature ($T\ge{}T_c$).
Data for S2 and S3 are shown in Supplemental Fig.~S1.
No evidence for a spontaneous Kerr rotation below T$_c$ is found for any of the investigated samples.

Under zero field conditions, prior measurements found that a spontaneous Kerr rotation developed below T$_c$ with amplitude less than or equal to 0.4~$\mu$rad ($\phi_0$)~\cite{Hayes2020,Wei2022}.
The Kerr rotation was also found to change sign and amplitude between different runs.
These changes were attributed to different chiral domain configurations that can form as the sample is cooled through T$_c$.
In the presence of random domains, the standard deviation of the spontaneous Kerr rotation between cooldowns ($\sigma_\textrm{S}$) is related to the ratio of the beam diameter ($w$) to the average domain size ($d$) \textit{via} the expression $\sigma_\textrm{S}=\phi{}_0d/w$ for $d<w$, wherein $\phi_0$ is the amplitude of the PKE arising from single domain~\cite{Xia2006b}.
The zero-field-cooled data presented by Wei \textit{et al.}~\cite{Wei2022} has a standard deviation ($\sigma$) of 0.097~$\mu$rad for T$>$T$_c$ and 0.244~$\mu$rad for T$<$T$_c$ across six runs.
Assuming that the noise (scatter) in the data ($\sigma_\textrm{N}$) and the magnitude of the spontaneous Kerr signal are independent:

\begin{gather}
    \begin{aligned}
	T>T_c&: \sigma_+^2=\sigma_{\textrm{N}}^2\\
	T<T_c&: \sigma_-^2=\sigma_{\textrm{N}}^2+\sigma_{\textrm{S}}^2
    \end{aligned}
\end{gather}

\noindent
This provides a value for $\sigma_\textrm{S}$ of 0.224~$\mu$rad, and an approximate domain size of 6~$\mu{}$m using the value of $\phi_0$ above and the stated beam diameter of 10~$\mu$m.
The beam diameter used in this study was also 10~$\mu$m in diameter.

The values of $\sigma_+$ and $\sigma_-$ were calculated for each sample in this study.
In contrast to prior reports~\cite{Hayes2020,Wei2022}, there is no significant difference between the standard deviation above and below T$_c$.
Using our normal-state deviation $\sigma_+$ and the value of $\sigma_\textrm{S}$ determined from Wei \textit{et al.}~\cite{Wei2022} would imply a standard deviation in the superconducting state $\sigma_-$ of 0.329 and 0.226~$\mu$rad in our data for samples S1 and S4, respectively.
Such an increase, illustrated by the $\pm{}2\sigma$ dashed gray lines in Fig.~\ref{fig1}c and d, is not observed.

\begin{figure*}[!ht]
	\includegraphics[width=\textwidth]{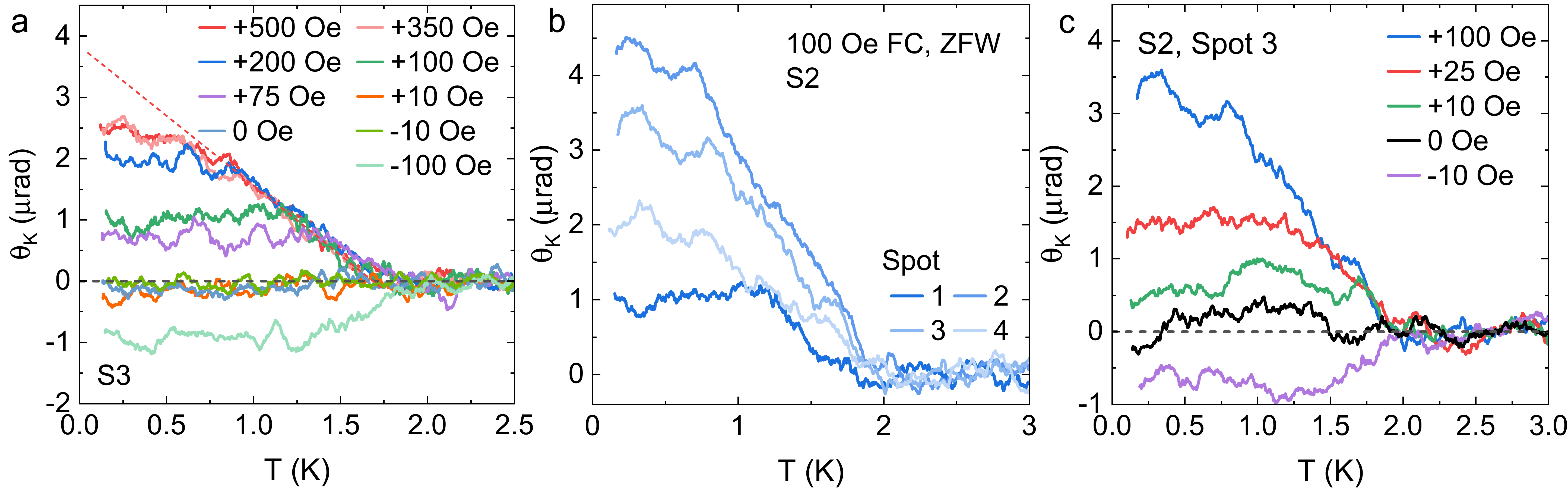}
    \caption{
		(a)~Zero-field-warmed $\theta_K$ after cooling in indicated field for spot 2 on S3.
		(b)~Zero-field-warmed $\theta_K$ after cooling in +100~Oe for four different spots on S2.
		(c)~Zero-field-warmed $\theta_K$ after cooling in indicated field for spot 3 on S2.
		(d)~Field sweep at 4~K (4.4~K for S2) after subtracting off diamagnetic contribution from optics ($\theta_{K\textrm{,optics}}$). Inset shows zoomed-in view of S2, Spot~3.
        In all cases, field was applied along the $c$ axis.
		}
    \label{fig2}
\end{figure*}

Although there is no spontaneous Kerr rotation below T$_c$, all samples show a field-trainable Kerr signal that vanishes at T$_c$.
Figure~\ref{fig2}a shows the temperature dependence of $\theta_K$ for S3 when warmed in zero field after being cooled through T$_c$ in small magnetic fields.
A finite Kerr signal clearly emerges for $T<T_c$, and the size of the signal is proportional to the cooling field for fields up to $\pm200$~Oe.
At higher fields, the signal begins to saturate as indicated by the identical Kerr rotation between the curves at 350 and 500~Oe.

In Wei \textit{et al.}~\cite{Wei2022}, $\theta_K$ was also found to be proportional to the cooling field.
In that study, however, $\theta_K$ was always $\pm0.4$~$\mu$rad for cooling fields between $15$~Oe and $30$~Oe in magnitude with the sign of $\theta_K$ 
dependent on the sign of the training field.
For fields lower than $\pm$15~Oe, $\theta_K$ was distributed between $-0.4$~$\mu$rad  and $+0.4$~$\mu$rad as noted above.
Thus, aside from the low-field behavior, our results are consistent with those reported in reference~\cite{Wei2022}.

Now we turn to signs of sample inhomogeneity revealed by PKE measurements.
Fig.~\ref{fig2}b shows the temperature evolution
of $\theta_K$ on four different spots of sample S2, upon warming up after being cooled in a field of +100~Oe.
As illustrated by Fig.~2(b), there is more than a factor of four change in the size of $\theta_K$ between the spots measured in sample S2.
This inhomogeneity strongly suggests an extrinsic origin for the trainable Kerr effect in UTe$_2$.
In addition, Figure~\ref{fig2}c shows $\theta_K$ for spot 3 on S2 when warmed in zero field after being cooled in small magnetic fields.
Even in small fields ($<\pm{}30$~Oe), $\theta_K$ is proportional to the training field, which is in contrast to the behavior previously observed
in samples that show a double transition in specific heat~\cite{Hayes2020,Wei2022}. As discussed below, inhomogeneity
and vortex pinning by defects in samples with a double transition may be responsible for this discrepancy.

To further correlate the PKE behavior with sample quality, we investigate vortex pinning effects through ac magnetic susceptibility measurements.
Figure~3a presents the real and the imaginary parts of the ac magnetic susceptibility $\chi'+i\chi''$ of four different UTe$_2$ single crystals with a range of T$_c$ values between 1.48~K and 2.1~K.
The specific heat data obtained on the same samples are shown in Fig. 3b for comparison.
The temperatures of the diamagnetic drop in $\chi'$ and the dissipation peak in $\chi''$ are both consistent with the specific heat data.
Samples S5, S6, and S2 exhibit a single thermodynamic transition, whereas sample S7 shows two clear transitions at 1.64~K and 1.48~K.
Note that sample S2 is the same sample S2 on which PKE data was presented above.
Once again, $\gamma^*$ decreases systematically for samples with higher T$_c$.
Although it has been previously argued that the double transition in some UTe$_2$ crystals is an intrinsic feature~\cite{Hayes2020}, it most often appears in crystals with lower T$_c$ or grown via CVT near the edge of the growth stability region for UTe$_2$~\cite{Weiland2022}.

\begin{figure}
	\includegraphics[width=\columnwidth]{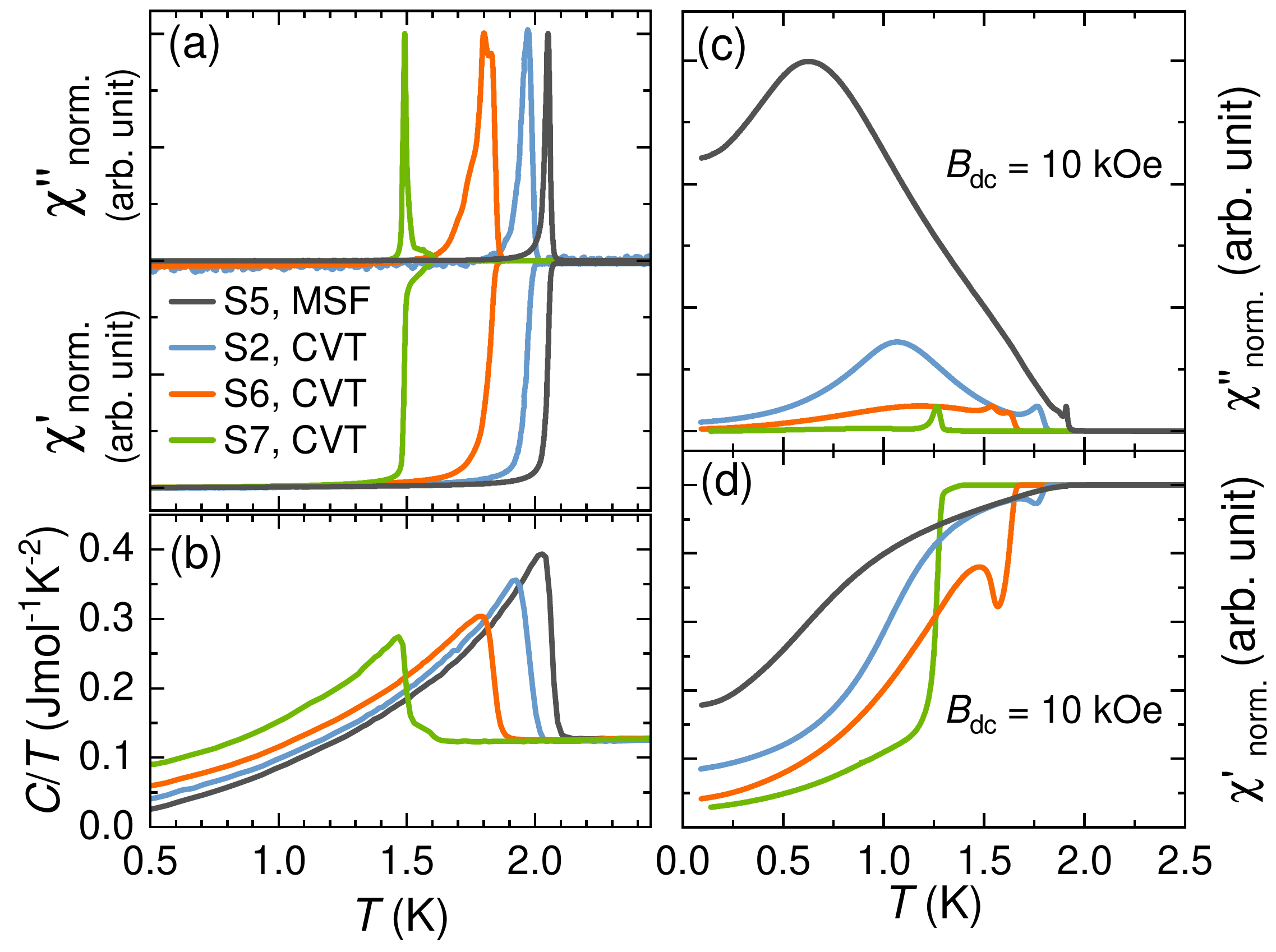}
    \caption{
		Specific heat and ac susceptibility on four samples with T$_c$ ranging from 1.48~K to 2.1~K.
		Sample S2 is the same sample on which Kerr effect was measured above.
		(a)~Normalized ac susceptibility versus temperature at zero magnetic field with an ac field of 0.25~Oe at 733~Hz.
		$\chi'$ represents the in-phase component and is normalized to a value of $-1$ at base temperature, whereas $\chi''$ is the out-of-phase component and is normalized to $+1$ at the maximum value.
		(b)~Specific heat versus temperature.
        (c) Normalized in-phase and (d) out-of-phase component of ac susceptibility cooled in 10~kOe applied field.
		$\chi'$ is normalized according to the measurement in zero field (panel a).
		$\chi''$ is normalized to $+1$ at the peak that indicates the onset of superconductivity.
		For both (c) and (d) an ac field of 2~Oe at 733~Hz was applied.
        All fields were applied along the $a$ axis.
		}
    \label{fig3}
\end{figure}

To investigate whether lower T$_c$ samples also have lower crystalline quality, Figs.~3c and 3d show ac susceptibility measurements cooled in a dc field of 10~kOe applied along the $a$ axis.
As T$_c$ increases, the relative drop in $\chi'$ below T$_c$ decreases for a fixed magnetic field.
To understand this apparent decrease in superconducting screening, we recall that, within the critical state model, the field H$^*$ at which magnetic flux first penetrates into the center of the sample after zero field cooling is proportional to the critical current (J$_c$) times the effective radius of the sample~\cite{Bean1964,Clem1993}.
In addition, lower defect density typically decreases J$_c$ and, as a result, H$^*$.
For a given field, more flux will therefore penetrate in a higher-quality sample compared to a lower-quality sample.
This is the reason for the enhanced $\chi'$ (reduced screening) at low temperatures in samples with higher T$_c$.
The relationship between vortex pinning strength and T$_c$ can be seen more clearly in the imaginary part of the susceptibility ($\chi''$), which reveals a peak effect characteristic of vortex lattice dynamics within the mixed state of type-II superconductors~\cite{Banerjee2011}.
A broad peak in $\chi''$ is observed below T$_c$ due to dissipation from vortex dynamics.
As vortex pinning becomes stronger, the broad peak in $\chi''$ decreases due to a decrease in dissipation.
Thus, the smaller peak in $\chi''$ is thereby evidence of stronger vortex pinning in lower T$_c$ samples.

\subsection{Discussion}

The absence of a spontaneous polar Kerr effect in our investigation of representative single-T$_c$ UTe$_2$ samples requires a careful discussion of consistency checks.
First, we note that the size of putative chiral domains should increase with increasing sample quality~\cite{Kallin2016a}.
In the case where the domain size is large compared to the beam diameter, $\sigma_\textrm{S}$ will equal $\phi_0$ (0.4~$\mu$rad).
Our ac magnetic susceptibility results demonstrate that higher T$_c$ samples have fewer defects, and 
the spontaneous polar Kerr response should therefore be even higher in the samples measured here if it were an intrinsic effect.

Second, we consider the effects of remnant magnetization in a type-II superconductor after cooling in constant field and removing the field at base temperature.
Clem and Hao~\cite{Clem1993} have calculated the expected remnant magnetization as a function of vortex pinning strength.
In a superconductor with strong vortex pinning, the remnant magnetization is expected to be nearly constant near the center of the sample and to decrease near the edges over a length scale determined by the vortex pinning strength.
Near the center, the remnant magnetization is expected to be non-zero up to the irreversibility temperature ($T_{\textrm{irr}}$), which has been shown to be very close to $T_c$ in UTe$_2$ for small magnetic fields~\cite{Rosuel2022a}.
Although remnant magnetization has been measured in UPt$_3$ and URu$_2$Si$_2$~\cite{Koziol1994}, prior measurements of $\theta_K$ in these systems have not found a signal that scales with the magnitude of the training field~\cite{Schemm2014,Schemm2015}.
It was argued that the reason for the difference between UTe$_2$ and other U-based materials is that in UTe$_2$ the vortices carry a ferromagnetic moment~\cite{Wei2022}.

If this were the case, and the observed Kerr effect in UTe$_2$ were due to intrinsic magnetism, one would expect that $\theta_K$ would be uniform across a large region in the center of the sample due to the relatively strong vortex pinning~\cite{Clem1993}.
Our position-dependent PKE measurements in sample S2, however, reveal a significant variation in the magnitude of $\theta_K$ near the center of the sample (see Fig.~\ref{fig2}b).
This spatial inhomogeneity strongly suggests an extrinsic origin for the field-trainable Kerr effect in UTe$_2$.
Notably, this sample has the lowest $\gamma^*$ of the three CVT-grown samples measured here.
Sample S4 also shows a large difference between measured spots (see Supplemental Fig.~S4).
If the origin of the finite Kerr signal were solely the inhomogenous magnetic clusters detected via $\mu$SR~\cite{Sundar2022}, one may expect a direct correlation between $\gamma^*$ and the size of the Kerr signal at a given training field.
Instead, the CVT-grown sample with the lowest $\gamma^*$ has the highest variability of $\theta_K$, which indicates the presence of another contribution to the field-trainable Kerr signal.
One obvious distinction is that $\mu$SR is a bulk probe, whereas PKE is surface sensitive.
From temperature-dependent optical conductivity measurements, the penetration depth for 1550~nm light is estimated to be only 300~nm~\cite{Mekonen2022}.
In addition, PKE probes a small surface area ($\approx$10~$\mu$m diameter) and will therefore depend on the local defect structure.
In fact, $\mu$SR measurements detect fluctuating magnetic clusters~\cite{Sundar2022}, and these fluctuations may be pinned by defects to become locally static~\cite{Iguchi2022}.
This may lead to a PKE signal that depends strongly on the proximity of the beam with respect to a pinning center.

Third, we note that the saturation of $\theta_K$ in the highest-quality sample (S1, T$_c$~=~2.1~K) occurs between 25 and 100 Oe,
whereas a higher field between 200 and 350 Oe  was required to saturate $\theta_K$ in sample S3 (T$_c$~=~1.9~K).
The field where saturation occurs is related to H$^*$, which is the field where the magnetic flux first penetrates into the center of the sample~\cite{Wei2022}.
At first sight, this appears to be consistent with the model of Clem and Hao, where H$^{*}$ is expected to decrease as the vortex pinning strength decreases~\cite{Clem1993}.
The remnant magnetization, however, is expected to scale with the vortex pinning strength.
If the magnitude of $\theta_K$ were related to the remnant magnetization of the bulk, then it should decrease as sample quality is improved.
Instead, it varies between spots on a given samples and between samples with no apparent trend.
This provides further evidence that the field-trainable Kerr effect is of extrinsic origin.

One significant difference between samples measured here and in prior reports is whether they host two transitions in specific heat.
As has been shown by several studies~\cite{Thomas2021,Rosa2022,Aoki2022}, the double transition is not an intrinsic feature of UTe$_2$.
Further, no evidence for a splitting of T$_c$ was found under shear uniaxial stress that would couple to the proposed $B_{3u}+iB_{2u}$ superconducting order parameter~\cite{Girod2022}.
In the face of our results, a plausible explanation for the difference in spontaneous Kerr signal between samples showing either a single or double transition 
is that the spatial inhomogeneity of the superconducting state may help nucleate and pin magnetic flux from fluctuating magnetic clusters and vortices.
In fact, extended crystallographic defects (e.g. line dislocations and grain boundaries) provide a natural explanation for stronger vortex pinning in lower T$_c$ samples, and
recent scanning SQUID measurements reveal that vortices tend to form along extended defects~\cite{Iguchi2022}.

Finally, we discuss the implications of our result for the superconducting order parameter of UTe$_2$. 
If the superconducting state in UTe$_2$ does not break time-reversal symmetry, proposals in favor of a chiral order parameter must be reevaluated.
Aside from PKE measurements~\cite{Hayes2020,Wei2022}, a chiral order parameter has also been suggested from the temperature dependence of magnetic penetration depth measurements of CVT-grown samples that indicated multiple pairs of point nodes~\cite{Ishihara2021}.
The samples investigated in Ref.~\cite{Ishihara2021} display a large residual specific heat ($\gamma^*\ge50$~mJ~mol$^{-1}$~K$^{-2}$), which indicates a large contribution from nonsuperconducting parts of the sample.
With the advent of molten salt flux growth~\cite{Sakai2022}, it is crucial to revisit prior experiments to determine 
which properties of UTe$_2$ are intrinsic and which are instead due to sample inhomogeneity, impurities, and crystallographic defects.
Other reports that support a chiral order parameter rely heavily on the observation of a TRSB order parameter to rule out other possibilities~\cite{Jiao2019,Bae2021}.
Further, no evidence for chiral domains or half-flux quantum vortices was observed in scanning SQUID measurements~\cite{Iguchi2022}.
We therefore must conclude that a chiral order parameter in UTe$_2$ is unlikely.

\subsection{Conclusion}

In summary, we performed systematic polar Kerr effect and ac magnetic susceptibility measurements on representative UTe$_2$ crystals of different quality to investigate the interplay between sample quality and time-reversal symmetry breaking.
Our results show that samples with a single thermodynamic superconducting transition do not develop a spontaneous Kerr rotation in the superconducting state.
In addition, the field-trainable Kerr effect observed in these samples is inhomogeneous and sample dependent, which strongly suggests an extrinsic origin.
We conclude that the superconducting state UTe$_2$ does not break time-reversal symmetry and a chiral order parameter is unlikely in UTe$_2$.

\subsection{Materials and Methods}
UTe$_2$ samples were grown via both chemical vapor transport ~\cite{Rosa2022} and molten salt flux~\cite{Sakai2022} techniques.
One noted drawback of the MSF technique is the potential inclusion of ferromagnetic U$_7$Te$_{12}$ with ordering temperature of 48~K~\cite{Opletal2022} or other uranium-tellerium binaries with ordering temperatures near 115~K~\cite{Tougait2001}.
In sample S1, which was grown by the MSF method, ferromagnetic impurities ($\theta_K\approx1$~mrad) were found on the surface, even after polishing, that tended to grow in thin strips along the $a$ axis.
The presence of ferromagnetic impurities was also confirmed by measurements of magnetic susceptibility.

Ac susceptibility was measured in a set of commercially wound compensated susceptibility coils.
Note that the data are normalized with respect to the height of the peak in $\chi''$ at T$_c$ for each sample.
This allows comparison between samples of different sizes.
The effectiveness of this normalization is shown in supplemental figure S6 where two pieces of sample S2 of different size were measured and seen to have similar behavior after this normalization.

Specific heat was measured using the quasi-adiabatic thermal relaxation technique in a He-3 cryostat insert.
Both small-pulse and long-pulse methods were used to ensure that all samples measured only had a single transition in specific heat.
Electrical resistivity was collected with a Lakeshore 372 AC bridge using a standard four-probe configuration wherein Pt wires were attached to sputtered gold
pads with silver paint. The residual resistivity ratio was determined by fitting the low-temperature resistivity to $\rho{\left(T\right)=AT^2+\rho_0}$ and taking the ratio $\rho{\left(300\textrm{~K}\right)}/\rho_0$.
Sample S2 was not measured directly, but the RRR for samples measured from the same growth batch was approximately 100.

PKE measurements were performed using a fiber-based zero area Sagnac interferometer operating at a wavelength of 1550~nm~\cite{Xia2006b,Fried2014b}.
Samples were mounted on a set of XYZ cryogenic piezo-stepper devices and thermally anchored to the base plate of an adiabatic demagnetization refrigerator.
Two different measurements types were performed: i) heat-pulse experiments to look for the development of a spontaneous PKE below T$_c$, and ii) field-cooled, zero-field-warmed measurements to investigate the field-trainable Kerr effect.

An example of the measurement process for the heat pulse experiments (i) is shown in Fig.~\ref{fig1}b.
To determine whether a spontaneous PKE onsets at T$_c$, samples were first cooled under zero magnetic field and illumination.
Maintaining zero-field conditions, a high intensity laser pulse was applied to the sample to heat the sample above T$_c$.
The left axis of Fig.~\ref{fig1}b shows V$_{2\omega}$, which is proportional to the light received by the photodetector.
The initial spike in V$_{2\omega}$ is caused by the high intensity laser pulse that is used to heat the sample.
It does not reach a very large value in the plot because of the short pulse time compared (7 seconds) to the settling time of the lockin amplifier (30 seconds).
Importantly, the resistivity becomes non-zero during the pulse indicating that the sample is heated back into the normal state.
The sample is then allowed to cool back to the current base temperature, which is the indicated temperature on Figs.~1c and 1d.
Next, the PKE of the sample was measured for several minutes at the current base temperature using 10--20~$\mu$W of optical power.
This is the region in Fig.~\ref{fig1}b where V$_{2\omega}$ is nearly 30~mV and the sample resistance remains zero.
This process was repeated as the base temperature of the system increased until the system warmed above T$_c$ to approximately 3~K.

In field-cooled, zero-field-warmed measurements (ii) the sample was cooled in a constant magnetic field.
After performing the demagnetization step needed to cool the system to base temperature, a the indicated field was applied to the sample.
A high intensity laser pulse was then used to temporarily heat the sample above T$_c$ without significantly raising the temperature of the base system.
During the pulse, the sample temperature rise above T$_c$ was confirmed by measuring the resistivity change in the sample.
This process ensured that the sample cooled through T$_c$ in the intended field and was not affected by the stray field from the demagnetization process.
After the single heat pulse the field was set to zero, and the sample was warmed at a fixed rate by controlling the base temperature of the system.
Measurements were typically made using 10--20~$\mu$W of incident optical power as the sample was warmed.
A small offset was subtracted from each curve so that the mean of $\theta_K$ was zero for $T>T_c$.
This offset is from a small remnant field ($<\pm5$~Oe) that arises from trapped flux in the demagnetization and sample space superconducting magnets.
Because the relationship between $\theta_K$ and the remnant field is known for $T>T_c$, it can be used to confirm the small magnitude of the remnant field.

Samples S1--S4 were prepared for optical measurements by polishing the surface.
The Supplemental Information also contains data on sample S8, which had as-grown surfaces that were suitable for optical measurements without polishing.
No notable differences were observed in the measurements between the as-grown and polished surfaces.
Additional heat pulse experiments were performed on S8 for which a local resistive heater was used to heat the sample above T$_c$ instead of the laser.
This allowed for the sample to cool from above T$_c$ back to base temperature over a period of 60~seconds instead of a few seconds.
No differences related to the cooling rate were observed.

\begin{acknowledgments}
Work at Los Alamos was supported by U.S. Department of Energy, Office of Basic Energy Sciences, Division of Materials Science and Engineering project ``Quantum Fluctuations in Narrow-Band Systems''.
Development of the Kerr effect measurements was supported by the Los Alamos Laboratory Directed Research and Development program.
MMB and BM acknowledge support from the Los Alamos Laboratory Directed Research and Development program.
\end{acknowledgments}

\bibliography{lib}

\begin{thebibliography}{35}
\expandafter\ifx\csname natexlab\endcsname\relax\def\natexlab#1{#1}\fi
\expandafter\ifx\csname bibnamefont\endcsname\relax
  \def\bibnamefont#1{#1}\fi
\expandafter\ifx\csname bibfnamefont\endcsname\relax
  \def\bibfnamefont#1{#1}\fi
\expandafter\ifx\csname citenamefont\endcsname\relax
  \def\citenamefont#1{#1}\fi
\expandafter\ifx\csname url\endcsname\relax
  \def\url#1{\texttt{#1}}\fi
\expandafter\ifx\csname urlprefix\endcsname\relax\def\urlprefix{URL }\fi
\providecommand{\bibinfo}[2]{#2}
\providecommand{\eprint}[2][]{\url{#2}}

\bibitem[{\citenamefont{Sigrist and Ueda}(1991)}]{Sigrist1991}
\bibinfo{author}{\bibfnamefont{M.}~\bibnamefont{Sigrist}} \bibnamefont{and}
  \bibinfo{author}{\bibfnamefont{K.}~\bibnamefont{Ueda}},
  \bibinfo{journal}{Reviews of Modern Physics} \textbf{\bibinfo{volume}{63}},
  \bibinfo{pages}{239} (\bibinfo{year}{1991}), ISSN \bibinfo{issn}{0034-6861},
  \urlprefix\url{https://link.aps.org/doi/10.1103/RevModPhys.63.239}.

\bibitem[{\citenamefont{Kallin and Berlinsky}(2016)}]{Kallin2016a}
\bibinfo{author}{\bibfnamefont{C.}~\bibnamefont{Kallin}} \bibnamefont{and}
  \bibinfo{author}{\bibfnamefont{J.}~\bibnamefont{Berlinsky}},
  \bibinfo{journal}{Reports on Progress in Physics}
  \textbf{\bibinfo{volume}{79}}, \bibinfo{pages}{054502}
  (\bibinfo{year}{2016}), ISSN \bibinfo{issn}{0034-4885}, \eprint{1512.01151},
  \urlprefix\url{https://iopscience.iop.org/article/10.1088/0034-4885/79/5/054502}.

\bibitem[{\citenamefont{Sato and Ando}(2017)}]{Sato2017a}
\bibinfo{author}{\bibfnamefont{M.}~\bibnamefont{Sato}} \bibnamefont{and}
  \bibinfo{author}{\bibfnamefont{Y.}~\bibnamefont{Ando}},
  \bibinfo{journal}{Reports on Progress in Physics}
  \textbf{\bibinfo{volume}{80}}, \bibinfo{pages}{076501}
  (\bibinfo{year}{2017}), ISSN \bibinfo{issn}{0034-4885}, \eprint{1608.03395},
  \urlprefix\url{http://stacks.iop.org/0034-4885/80/i=7/a=076501?key=crossref.db5f5db067ad36cf0669f5cbf8ccb916}.

\bibitem[{\citenamefont{Mackenzie et~al.}(2017)\citenamefont{Mackenzie,
  Scaffidi, Hicks, and Maeno}}]{Mackenzie2017}
\bibinfo{author}{\bibfnamefont{A.~P.} \bibnamefont{Mackenzie}},
  \bibinfo{author}{\bibfnamefont{T.}~\bibnamefont{Scaffidi}},
  \bibinfo{author}{\bibfnamefont{C.~W.} \bibnamefont{Hicks}}, \bibnamefont{and}
  \bibinfo{author}{\bibfnamefont{Y.}~\bibnamefont{Maeno}},
  \bibinfo{journal}{npj Quantum Materials} \textbf{\bibinfo{volume}{2}},
  \bibinfo{pages}{40} (\bibinfo{year}{2017}), ISSN \bibinfo{issn}{2397-4648},
  \eprint{1706.01942},
  \urlprefix\url{http://www.nature.com/articles/s41535-017-0045-4}.

\bibitem[{\citenamefont{Pustogow et~al.}(2019)\citenamefont{Pustogow, Luo,
  Chronister, Su, Sokolov, Jerzembeck, Mackenzie, Hicks, Kikugawa, Raghu
  et~al.}}]{Pustogow2019}
\bibinfo{author}{\bibfnamefont{A.}~\bibnamefont{Pustogow}},
  \bibinfo{author}{\bibfnamefont{Y.}~\bibnamefont{Luo}},
  \bibinfo{author}{\bibfnamefont{A.}~\bibnamefont{Chronister}},
  \bibinfo{author}{\bibfnamefont{Y.~S.} \bibnamefont{Su}},
  \bibinfo{author}{\bibfnamefont{D.~A.} \bibnamefont{Sokolov}},
  \bibinfo{author}{\bibfnamefont{F.}~\bibnamefont{Jerzembeck}},
  \bibinfo{author}{\bibfnamefont{A.~P.} \bibnamefont{Mackenzie}},
  \bibinfo{author}{\bibfnamefont{C.~W.} \bibnamefont{Hicks}},
  \bibinfo{author}{\bibfnamefont{N.}~\bibnamefont{Kikugawa}},
  \bibinfo{author}{\bibfnamefont{S.}~\bibnamefont{Raghu}},
  \bibnamefont{et~al.}, \bibinfo{journal}{Nature}
  \textbf{\bibinfo{volume}{574}}, \bibinfo{pages}{72} (\bibinfo{year}{2019}),
  ISSN \bibinfo{issn}{14764687}.

\bibitem[{\citenamefont{Ran et~al.}(2019)\citenamefont{Ran, Eckberg, Ding,
  Furukawa, Metz, Saha, Liu, Zic, Kim, Paglione et~al.}}]{Ran2019}
\bibinfo{author}{\bibfnamefont{S.}~\bibnamefont{Ran}},
  \bibinfo{author}{\bibfnamefont{C.}~\bibnamefont{Eckberg}},
  \bibinfo{author}{\bibfnamefont{Q.-P.} \bibnamefont{Ding}},
  \bibinfo{author}{\bibfnamefont{Y.}~\bibnamefont{Furukawa}},
  \bibinfo{author}{\bibfnamefont{T.}~\bibnamefont{Metz}},
  \bibinfo{author}{\bibfnamefont{S.~R.} \bibnamefont{Saha}},
  \bibinfo{author}{\bibfnamefont{I.-L.} \bibnamefont{Liu}},
  \bibinfo{author}{\bibfnamefont{M.}~\bibnamefont{Zic}},
  \bibinfo{author}{\bibfnamefont{H.}~\bibnamefont{Kim}},
  \bibinfo{author}{\bibfnamefont{J.}~\bibnamefont{Paglione}},
  \bibnamefont{et~al.}, \bibinfo{journal}{Science}
  \textbf{\bibinfo{volume}{365}}, \bibinfo{pages}{684} (\bibinfo{year}{2019}),
  ISSN \bibinfo{issn}{0036-8075},
  \urlprefix\url{http://www.sciencemag.org/lookup/doi/10.1126/science.aav8645}.

\bibitem[{\citenamefont{Jiao et~al.}(2020)\citenamefont{Jiao, Howard, Ran,
  Wang, Rodriguez, Sigrist, Wang, Butch, and Madhavan}}]{Jiao2019}
\bibinfo{author}{\bibfnamefont{L.}~\bibnamefont{Jiao}},
  \bibinfo{author}{\bibfnamefont{S.}~\bibnamefont{Howard}},
  \bibinfo{author}{\bibfnamefont{S.}~\bibnamefont{Ran}},
  \bibinfo{author}{\bibfnamefont{Z.}~\bibnamefont{Wang}},
  \bibinfo{author}{\bibfnamefont{J.~O.} \bibnamefont{Rodriguez}},
  \bibinfo{author}{\bibfnamefont{M.}~\bibnamefont{Sigrist}},
  \bibinfo{author}{\bibfnamefont{Z.}~\bibnamefont{Wang}},
  \bibinfo{author}{\bibfnamefont{N.~P.} \bibnamefont{Butch}}, \bibnamefont{and}
  \bibinfo{author}{\bibfnamefont{V.}~\bibnamefont{Madhavan}},
  \bibinfo{journal}{Nature} \textbf{\bibinfo{volume}{579}},
  \bibinfo{pages}{523} (\bibinfo{year}{2020}), ISSN \bibinfo{issn}{0028-0836},
  \eprint{1908.02846}, \urlprefix\url{http://arxiv.org/abs/1908.02846
  http://dx.doi.org/10.1038/s41586-020-2122-2
  http://www.nature.com/articles/s41586-020-2122-2}.

\bibitem[{\citenamefont{Bae et~al.}(2021)\citenamefont{Bae, Kim, Eo, Ran, Liu,
  Fuhrman, Paglione, Butch, and Anlage}}]{Bae2021}
\bibinfo{author}{\bibfnamefont{S.}~\bibnamefont{Bae}},
  \bibinfo{author}{\bibfnamefont{H.}~\bibnamefont{Kim}},
  \bibinfo{author}{\bibfnamefont{Y.~S.} \bibnamefont{Eo}},
  \bibinfo{author}{\bibfnamefont{S.}~\bibnamefont{Ran}},
  \bibinfo{author}{\bibfnamefont{I.-l.} \bibnamefont{Liu}},
  \bibinfo{author}{\bibfnamefont{W.~T.} \bibnamefont{Fuhrman}},
  \bibinfo{author}{\bibfnamefont{J.}~\bibnamefont{Paglione}},
  \bibinfo{author}{\bibfnamefont{N.~P.} \bibnamefont{Butch}}, \bibnamefont{and}
  \bibinfo{author}{\bibfnamefont{S.~M.} \bibnamefont{Anlage}},
  \bibinfo{journal}{Nature Communications} \textbf{\bibinfo{volume}{12}},
  \bibinfo{pages}{2644} (\bibinfo{year}{2021}), ISSN \bibinfo{issn}{2041-1723},
  \eprint{1909.09032},
  \urlprefix\url{https://www.nature.com/articles/s41467-021-22906-6}.

\bibitem[{\citenamefont{Ishihara et~al.}(2021)\citenamefont{Ishihara, Roppongi,
  Kobayashi, Mizukami, Sakai, Haga, Hashimoto, and Shibauchi}}]{Ishihara2021}
\bibinfo{author}{\bibfnamefont{K.}~\bibnamefont{Ishihara}},
  \bibinfo{author}{\bibfnamefont{M.}~\bibnamefont{Roppongi}},
  \bibinfo{author}{\bibfnamefont{M.}~\bibnamefont{Kobayashi}},
  \bibinfo{author}{\bibfnamefont{Y.}~\bibnamefont{Mizukami}},
  \bibinfo{author}{\bibfnamefont{H.}~\bibnamefont{Sakai}},
  \bibinfo{author}{\bibfnamefont{Y.}~\bibnamefont{Haga}},
  \bibinfo{author}{\bibfnamefont{K.}~\bibnamefont{Hashimoto}},
  \bibnamefont{and}
  \bibinfo{author}{\bibfnamefont{T.}~\bibnamefont{Shibauchi}},
  \bibinfo{journal}{arXiv} pp. \bibinfo{pages}{1--32} (\bibinfo{year}{2021}),
  \eprint{2105.13721}, \urlprefix\url{http://arxiv.org/abs/2105.13721}.

\bibitem[{\citenamefont{Joynt and Taillefer}(2002)}]{Joynt2002}
\bibinfo{author}{\bibfnamefont{R.}~\bibnamefont{Joynt}} \bibnamefont{and}
  \bibinfo{author}{\bibfnamefont{L.}~\bibnamefont{Taillefer}},
  \bibinfo{journal}{Reviews of Modern Physics} \textbf{\bibinfo{volume}{74}},
  \bibinfo{pages}{235} (\bibinfo{year}{2002}), ISSN \bibinfo{issn}{00346861}.

\bibitem[{\citenamefont{Schemm et~al.}(2014)\citenamefont{Schemm, Gannon,
  Wishne, Halperin, and Kapitulnik}}]{Schemm2014}
\bibinfo{author}{\bibfnamefont{E.~R.} \bibnamefont{Schemm}},
  \bibinfo{author}{\bibfnamefont{W.~J.} \bibnamefont{Gannon}},
  \bibinfo{author}{\bibfnamefont{C.~M.} \bibnamefont{Wishne}},
  \bibinfo{author}{\bibfnamefont{W.~P.} \bibnamefont{Halperin}},
  \bibnamefont{and}
  \bibinfo{author}{\bibfnamefont{A.}~\bibnamefont{Kapitulnik}},
  \bibinfo{journal}{Science} \textbf{\bibinfo{volume}{345}},
  \bibinfo{pages}{190} (\bibinfo{year}{2014}), ISSN \bibinfo{issn}{0036-8075},
  \urlprefix\url{https://www.sciencemag.org/lookup/doi/10.1126/science.1248552}.

\bibitem[{\citenamefont{Avers et~al.}(2020)\citenamefont{Avers, Gannon, Kuhn,
  Halperin, Sauls, DeBeer-Schmitt, Dewhurst, Gavilano, Nagy, Gasser
  et~al.}}]{Avers2020}
\bibinfo{author}{\bibfnamefont{K.~E.} \bibnamefont{Avers}},
  \bibinfo{author}{\bibfnamefont{W.~J.} \bibnamefont{Gannon}},
  \bibinfo{author}{\bibfnamefont{S.~J.} \bibnamefont{Kuhn}},
  \bibinfo{author}{\bibfnamefont{W.~P.} \bibnamefont{Halperin}},
  \bibinfo{author}{\bibfnamefont{J.~A.} \bibnamefont{Sauls}},
  \bibinfo{author}{\bibfnamefont{L.}~\bibnamefont{DeBeer-Schmitt}},
  \bibinfo{author}{\bibfnamefont{C.~D.} \bibnamefont{Dewhurst}},
  \bibinfo{author}{\bibfnamefont{J.}~\bibnamefont{Gavilano}},
  \bibinfo{author}{\bibfnamefont{G.}~\bibnamefont{Nagy}},
  \bibinfo{author}{\bibfnamefont{U.}~\bibnamefont{Gasser}},
  \bibnamefont{et~al.}, \bibinfo{journal}{Nature Physics}
  (\bibinfo{year}{2020}), ISSN \bibinfo{issn}{1745-2473},
  \urlprefix\url{http://dx.doi.org/10.1038/s41567-020-0822-z
  http://www.nature.com/articles/s41567-020-0822-z}.

\bibitem[{\citenamefont{Hayes et~al.}(2021)\citenamefont{Hayes, Wei, Metz,
  Zhang, Eo, Ran, Saha, Collini, Butch, Agterberg et~al.}}]{Hayes2020}
\bibinfo{author}{\bibfnamefont{I.~M.} \bibnamefont{Hayes}},
  \bibinfo{author}{\bibfnamefont{D.~S.} \bibnamefont{Wei}},
  \bibinfo{author}{\bibfnamefont{T.}~\bibnamefont{Metz}},
  \bibinfo{author}{\bibfnamefont{J.}~\bibnamefont{Zhang}},
  \bibinfo{author}{\bibfnamefont{Y.~S.} \bibnamefont{Eo}},
  \bibinfo{author}{\bibfnamefont{S.}~\bibnamefont{Ran}},
  \bibinfo{author}{\bibfnamefont{S.~R.} \bibnamefont{Saha}},
  \bibinfo{author}{\bibfnamefont{J.}~\bibnamefont{Collini}},
  \bibinfo{author}{\bibfnamefont{N.~P.} \bibnamefont{Butch}},
  \bibinfo{author}{\bibfnamefont{D.~F.} \bibnamefont{Agterberg}},
  \bibnamefont{et~al.}, \bibinfo{journal}{Science}
  \textbf{\bibinfo{volume}{373}}, \bibinfo{pages}{797} (\bibinfo{year}{2021}),
  ISSN \bibinfo{issn}{0036-8075}, \eprint{2002.02539},
  \urlprefix\url{http://arxiv.org/abs/2002.02539
  https://www.sciencemag.org/lookup/doi/10.1126/science.abb0272}.

\bibitem[{\citenamefont{Wei et~al.}(2022)\citenamefont{Wei, Saykin, Miller,
  Ran, Saha, Agterberg, Schmalian, Butch, Paglione, and Kapitulnik}}]{Wei2022}
\bibinfo{author}{\bibfnamefont{D.~S.} \bibnamefont{Wei}},
  \bibinfo{author}{\bibfnamefont{D.}~\bibnamefont{Saykin}},
  \bibinfo{author}{\bibfnamefont{O.~Y.} \bibnamefont{Miller}},
  \bibinfo{author}{\bibfnamefont{S.}~\bibnamefont{Ran}},
  \bibinfo{author}{\bibfnamefont{S.~R.} \bibnamefont{Saha}},
  \bibinfo{author}{\bibfnamefont{D.~F.} \bibnamefont{Agterberg}},
  \bibinfo{author}{\bibfnamefont{J.}~\bibnamefont{Schmalian}},
  \bibinfo{author}{\bibfnamefont{N.~P.} \bibnamefont{Butch}},
  \bibinfo{author}{\bibfnamefont{J.}~\bibnamefont{Paglione}}, \bibnamefont{and}
  \bibinfo{author}{\bibfnamefont{A.}~\bibnamefont{Kapitulnik}},
  \bibinfo{journal}{Physical Review B} \textbf{\bibinfo{volume}{105}},
  \bibinfo{pages}{024521} (\bibinfo{year}{2022}), ISSN
  \bibinfo{issn}{2469-9950}, \eprint{2108.09838},
  \urlprefix\url{https://link.aps.org/doi/10.1103/PhysRevB.105.024521}.

\bibitem[{\citenamefont{Rosa et~al.}(2022)\citenamefont{Rosa, Weiland, Fender,
  Scott, Ronning, Thompson, Bauer, and Thomas}}]{Rosa2022}
\bibinfo{author}{\bibfnamefont{P.~F.~S.} \bibnamefont{Rosa}},
  \bibinfo{author}{\bibfnamefont{A.}~\bibnamefont{Weiland}},
  \bibinfo{author}{\bibfnamefont{S.~S.} \bibnamefont{Fender}},
  \bibinfo{author}{\bibfnamefont{B.~L.} \bibnamefont{Scott}},
  \bibinfo{author}{\bibfnamefont{F.}~\bibnamefont{Ronning}},
  \bibinfo{author}{\bibfnamefont{J.~D.} \bibnamefont{Thompson}},
  \bibinfo{author}{\bibfnamefont{E.~D.} \bibnamefont{Bauer}}, \bibnamefont{and}
  \bibinfo{author}{\bibfnamefont{S.~M.} \bibnamefont{Thomas}},
  \bibinfo{journal}{Communications Materials} \textbf{\bibinfo{volume}{3}},
  \bibinfo{pages}{33} (\bibinfo{year}{2022}), ISSN \bibinfo{issn}{2662-4443},
  \urlprefix\url{https://www.nature.com/articles/s43246-022-00254-2}.

\bibitem[{\citenamefont{Weiland et~al.}(2022)\citenamefont{Weiland, Thomas, and
  Rosa}}]{Weiland2022}
\bibinfo{author}{\bibfnamefont{A.}~\bibnamefont{Weiland}},
  \bibinfo{author}{\bibfnamefont{S.~M.} \bibnamefont{Thomas}},
  \bibnamefont{and} \bibinfo{author}{\bibfnamefont{P.~F.~S.}
  \bibnamefont{Rosa}}, \bibinfo{journal}{Journal of Physics: Materials}
  \textbf{\bibinfo{volume}{5}}, \bibinfo{pages}{044001} (\bibinfo{year}{2022}),
  ISSN \bibinfo{issn}{2515-7639},
  \urlprefix\url{https://iopscience.iop.org/article/10.1088/2515-7639/ac8ba9}.

\bibitem[{\citenamefont{Thomas et~al.}(2021)\citenamefont{Thomas, Stevens,
  Santos, Fender, Bauer, Ronning, Thompson, Huxley, and Rosa}}]{Thomas2021}
\bibinfo{author}{\bibfnamefont{S.~M.} \bibnamefont{Thomas}},
  \bibinfo{author}{\bibfnamefont{C.}~\bibnamefont{Stevens}},
  \bibinfo{author}{\bibfnamefont{F.~B.} \bibnamefont{Santos}},
  \bibinfo{author}{\bibfnamefont{S.~S.} \bibnamefont{Fender}},
  \bibinfo{author}{\bibfnamefont{E.~D.} \bibnamefont{Bauer}},
  \bibinfo{author}{\bibfnamefont{F.}~\bibnamefont{Ronning}},
  \bibinfo{author}{\bibfnamefont{J.~D.} \bibnamefont{Thompson}},
  \bibinfo{author}{\bibfnamefont{A.}~\bibnamefont{Huxley}}, \bibnamefont{and}
  \bibinfo{author}{\bibfnamefont{P.~F.~S.} \bibnamefont{Rosa}},
  \bibinfo{journal}{arXiv}  (\bibinfo{year}{2021}), \eprint{2103.09194},
  \urlprefix\url{http://arxiv.org/abs/2103.09194}.

\bibitem[{\citenamefont{Aoki et~al.}(2022{\natexlab{a}})\citenamefont{Aoki,
  Brison, Flouquet, Ishida, Knebel, Tokunaga, and Yanase}}]{Aoki2022}
\bibinfo{author}{\bibfnamefont{D.}~\bibnamefont{Aoki}},
  \bibinfo{author}{\bibfnamefont{J.-P.} \bibnamefont{Brison}},
  \bibinfo{author}{\bibfnamefont{J.}~\bibnamefont{Flouquet}},
  \bibinfo{author}{\bibfnamefont{K.}~\bibnamefont{Ishida}},
  \bibinfo{author}{\bibfnamefont{G.}~\bibnamefont{Knebel}},
  \bibinfo{author}{\bibfnamefont{Y.}~\bibnamefont{Tokunaga}}, \bibnamefont{and}
  \bibinfo{author}{\bibfnamefont{Y.}~\bibnamefont{Yanase}},
  \bibinfo{journal}{Journal of Physics: Condensed Matter}
  \textbf{\bibinfo{volume}{34}}, \bibinfo{pages}{243002}
  (\bibinfo{year}{2022}{\natexlab{a}}), ISSN \bibinfo{issn}{0953-8984},
  \urlprefix\url{https://iopscience.iop.org/article/10.1088/1361-648X/ac5863}.

\bibitem[{\citenamefont{Girod et~al.}(2022)\citenamefont{Girod, Stevens,
  Huxley, Bauer, Santos, Thompson, Fernandes, Zhu, Ronning, Rosa
  et~al.}}]{Girod2022}
\bibinfo{author}{\bibfnamefont{C.}~\bibnamefont{Girod}},
  \bibinfo{author}{\bibfnamefont{C.~R.} \bibnamefont{Stevens}},
  \bibinfo{author}{\bibfnamefont{A.}~\bibnamefont{Huxley}},
  \bibinfo{author}{\bibfnamefont{E.~D.} \bibnamefont{Bauer}},
  \bibinfo{author}{\bibfnamefont{F.~B.} \bibnamefont{Santos}},
  \bibinfo{author}{\bibfnamefont{J.~D.} \bibnamefont{Thompson}},
  \bibinfo{author}{\bibfnamefont{R.~M.} \bibnamefont{Fernandes}},
  \bibinfo{author}{\bibfnamefont{J.-x.} \bibnamefont{Zhu}},
  \bibinfo{author}{\bibfnamefont{F.}~\bibnamefont{Ronning}},
  \bibinfo{author}{\bibfnamefont{P.~F.~S.} \bibnamefont{Rosa}},
  \bibnamefont{et~al.}, \bibinfo{journal}{Physical Review B}
  \textbf{\bibinfo{volume}{106}}, \bibinfo{pages}{L121101}
  (\bibinfo{year}{2022}), ISSN \bibinfo{issn}{2469-9950},
  \urlprefix\url{https://link.aps.org/doi/10.1103/PhysRevB.106.L121101}.

\bibitem[{\citenamefont{Sakai et~al.}(2022)\citenamefont{Sakai, Opletal,
  Tokiwa, Yamamoto, Tokunaga, Kambe, and Haga}}]{Sakai2022}
\bibinfo{author}{\bibfnamefont{H.}~\bibnamefont{Sakai}},
  \bibinfo{author}{\bibfnamefont{P.}~\bibnamefont{Opletal}},
  \bibinfo{author}{\bibfnamefont{Y.}~\bibnamefont{Tokiwa}},
  \bibinfo{author}{\bibfnamefont{E.}~\bibnamefont{Yamamoto}},
  \bibinfo{author}{\bibfnamefont{Y.}~\bibnamefont{Tokunaga}},
  \bibinfo{author}{\bibfnamefont{S.}~\bibnamefont{Kambe}}, \bibnamefont{and}
  \bibinfo{author}{\bibfnamefont{Y.}~\bibnamefont{Haga}},
  \bibinfo{journal}{Physical Review Materials} \textbf{\bibinfo{volume}{6}},
  \bibinfo{pages}{073401} (\bibinfo{year}{2022}), ISSN
  \bibinfo{issn}{2475-9953},
  \urlprefix\url{https://link.aps.org/doi/10.1103/PhysRevMaterials.6.073401}.

\bibitem[{\citenamefont{Aoki et~al.}(2022{\natexlab{b}})\citenamefont{Aoki,
  Sakai, Opletal, Tokiwa, Ishizuka, Yanase, Harima, Nakamura, Li, Homma
  et~al.}}]{Aoki2022b}
\bibinfo{author}{\bibfnamefont{D.}~\bibnamefont{Aoki}},
  \bibinfo{author}{\bibfnamefont{H.}~\bibnamefont{Sakai}},
  \bibinfo{author}{\bibfnamefont{P.}~\bibnamefont{Opletal}},
  \bibinfo{author}{\bibfnamefont{Y.}~\bibnamefont{Tokiwa}},
  \bibinfo{author}{\bibfnamefont{J.}~\bibnamefont{Ishizuka}},
  \bibinfo{author}{\bibfnamefont{Y.}~\bibnamefont{Yanase}},
  \bibinfo{author}{\bibfnamefont{H.}~\bibnamefont{Harima}},
  \bibinfo{author}{\bibfnamefont{A.}~\bibnamefont{Nakamura}},
  \bibinfo{author}{\bibfnamefont{D.}~\bibnamefont{Li}},
  \bibinfo{author}{\bibfnamefont{Y.}~\bibnamefont{Homma}},
  \bibnamefont{et~al.}, \bibinfo{journal}{Journal of the Physical Society of
  Japan} \textbf{\bibinfo{volume}{91}}, \bibinfo{pages}{1}
  (\bibinfo{year}{2022}{\natexlab{b}}), ISSN \bibinfo{issn}{13474073},
  \eprint{2206.01363},
  \urlprefix\url{https://journals.jps.jp/doi/10.7566/JPSJ.91.083704}.

\bibitem[{\citenamefont{Cairns et~al.}(2020)\citenamefont{Cairns, Stevens,
  O'Neill, and Huxley}}]{Cairns2020}
\bibinfo{author}{\bibfnamefont{L.~P.} \bibnamefont{Cairns}},
  \bibinfo{author}{\bibfnamefont{C.~R.} \bibnamefont{Stevens}},
  \bibinfo{author}{\bibfnamefont{C.~D.} \bibnamefont{O'Neill}},
  \bibnamefont{and} \bibinfo{author}{\bibfnamefont{A.}~\bibnamefont{Huxley}},
  \bibinfo{journal}{Journal of Physics: Condensed Matter}
  \textbf{\bibinfo{volume}{32}}, \bibinfo{pages}{415602}
  (\bibinfo{year}{2020}), ISSN \bibinfo{issn}{0953-8984},
  \urlprefix\url{https://iopscience.iop.org/article/10.1088/1361-648X/ab9c5d}.

\bibitem[{\citenamefont{Sundar et~al.}(2022)\citenamefont{Sundar, Azari, Goeks,
  Gheidi, Abedi, Yakovlev, Dunsiger, Wilkinson, Blundell, Metz
  et~al.}}]{Sundar2022}
\bibinfo{author}{\bibfnamefont{S.}~\bibnamefont{Sundar}},
  \bibinfo{author}{\bibfnamefont{N.}~\bibnamefont{Azari}},
  \bibinfo{author}{\bibfnamefont{M.}~\bibnamefont{Goeks}},
  \bibinfo{author}{\bibfnamefont{S.}~\bibnamefont{Gheidi}},
  \bibinfo{author}{\bibfnamefont{M.}~\bibnamefont{Abedi}},
  \bibinfo{author}{\bibfnamefont{M.}~\bibnamefont{Yakovlev}},
  \bibinfo{author}{\bibfnamefont{S.~R.} \bibnamefont{Dunsiger}},
  \bibinfo{author}{\bibfnamefont{J.~M.} \bibnamefont{Wilkinson}},
  \bibinfo{author}{\bibfnamefont{S.~J.} \bibnamefont{Blundell}},
  \bibinfo{author}{\bibfnamefont{T.~E.} \bibnamefont{Metz}},
  \bibnamefont{et~al.}, pp. \bibinfo{pages}{1--33} (\bibinfo{year}{2022}),
  \eprint{2207.13725}, \urlprefix\url{http://arxiv.org/abs/2207.13725}.

\bibitem[{\citenamefont{Xia et~al.}(2006)\citenamefont{Xia, Beyersdorf, Fejer,
  and Kapitulnik}}]{Xia2006b}
\bibinfo{author}{\bibfnamefont{J.}~\bibnamefont{Xia}},
  \bibinfo{author}{\bibfnamefont{P.~T.} \bibnamefont{Beyersdorf}},
  \bibinfo{author}{\bibfnamefont{M.~M.} \bibnamefont{Fejer}}, \bibnamefont{and}
  \bibinfo{author}{\bibfnamefont{A.}~\bibnamefont{Kapitulnik}},
  \bibinfo{journal}{Applied Physics Letters} \textbf{\bibinfo{volume}{89}},
  \bibinfo{pages}{062508} (\bibinfo{year}{2006}), ISSN
  \bibinfo{issn}{0003-6951},
  \urlprefix\url{http://aip.scitation.org/doi/10.1063/1.2336620}.

\bibitem[{\citenamefont{Bean}(1964)}]{Bean1964}
\bibinfo{author}{\bibfnamefont{C.~P.} \bibnamefont{Bean}},
  \bibinfo{journal}{Reviews of Modern Physics} \textbf{\bibinfo{volume}{36}},
  \bibinfo{pages}{31} (\bibinfo{year}{1964}), ISSN \bibinfo{issn}{0034-6861},
  \urlprefix\url{http://aip.scitation.org/doi/10.1063/1.1713463
  https://link.aps.org/doi/10.1103/RevModPhys.36.31}.

\bibitem[{\citenamefont{Clem and Hao}(1993)}]{Clem1993}
\bibinfo{author}{\bibfnamefont{J.~R.} \bibnamefont{Clem}} \bibnamefont{and}
  \bibinfo{author}{\bibfnamefont{Z.}~\bibnamefont{Hao}},
  \bibinfo{journal}{Physical Review B} \textbf{\bibinfo{volume}{48}},
  \bibinfo{pages}{13774} (\bibinfo{year}{1993}), ISSN
  \bibinfo{issn}{0163-1829},
  \urlprefix\url{https://link.aps.org/doi/10.1103/PhysRevB.48.13774}.

\bibitem[{\citenamefont{Banerjee}(2011)}]{Banerjee2011}
\bibinfo{author}{\bibfnamefont{S.}~\bibnamefont{Banerjee}}, in
  \emph{\bibinfo{booktitle}{Superconductivity - Theory and Applications}},
  edited by \bibinfo{editor}{\bibfnamefont{A.}~\bibnamefont{{Moyses Luiz}}}
  (\bibinfo{publisher}{IntechOpen}, \bibinfo{year}{2011}),
  chap.~\bibinfo{chapter}{4}, pp. \bibinfo{pages}{55--84}, ISBN
  \bibinfo{isbn}{9789533071510}.

\bibitem[{\citenamefont{Rosuel et~al.}(2022)\citenamefont{Rosuel, Marcenat,
  Knebel, Klein, Pourret, Marquardt, Niu, Rousseau, Demuer, Seyfarth
  et~al.}}]{Rosuel2022a}
\bibinfo{author}{\bibfnamefont{A.}~\bibnamefont{Rosuel}},
  \bibinfo{author}{\bibfnamefont{C.}~\bibnamefont{Marcenat}},
  \bibinfo{author}{\bibfnamefont{G.}~\bibnamefont{Knebel}},
  \bibinfo{author}{\bibfnamefont{T.}~\bibnamefont{Klein}},
  \bibinfo{author}{\bibfnamefont{A.}~\bibnamefont{Pourret}},
  \bibinfo{author}{\bibfnamefont{N.}~\bibnamefont{Marquardt}},
  \bibinfo{author}{\bibfnamefont{Q.}~\bibnamefont{Niu}},
  \bibinfo{author}{\bibfnamefont{S.}~\bibnamefont{Rousseau}},
  \bibinfo{author}{\bibfnamefont{A.}~\bibnamefont{Demuer}},
  \bibinfo{author}{\bibfnamefont{G.}~\bibnamefont{Seyfarth}},
  \bibnamefont{et~al.}, pp. \bibinfo{pages}{1--26} (\bibinfo{year}{2022}),
  \eprint{2205.04524}, \urlprefix\url{http://arxiv.org/abs/2205.04524}.

\bibitem[{\citenamefont{Koziol et~al.}(1994)\citenamefont{Koziol, Franse,
  de~Chatel, and Menovsky}}]{Koziol1994}
\bibinfo{author}{\bibfnamefont{Z.}~\bibnamefont{Koziol}},
  \bibinfo{author}{\bibfnamefont{J.~J.~M.} \bibnamefont{Franse}},
  \bibinfo{author}{\bibfnamefont{P.~F.} \bibnamefont{de~Chatel}},
  \bibnamefont{and} \bibinfo{author}{\bibfnamefont{A.~A.}
  \bibnamefont{Menovsky}}, \bibinfo{journal}{Phys. Rev. B}
  \textbf{\bibinfo{volume}{50}}, \bibinfo{pages}{15978} (\bibinfo{year}{1994}),
  ISSN \bibinfo{issn}{0163-1829},
  \urlprefix\url{https://link.aps.org/doi/10.1103/PhysRevB.50.15978}.

\bibitem[{\citenamefont{Schemm et~al.}(2015)\citenamefont{Schemm, Baumbach,
  Tobash, Ronning, Bauer, and Kapitulnik}}]{Schemm2015}
\bibinfo{author}{\bibfnamefont{E.~R.} \bibnamefont{Schemm}},
  \bibinfo{author}{\bibfnamefont{R.~E.} \bibnamefont{Baumbach}},
  \bibinfo{author}{\bibfnamefont{P.~H.} \bibnamefont{Tobash}},
  \bibinfo{author}{\bibfnamefont{F.}~\bibnamefont{Ronning}},
  \bibinfo{author}{\bibfnamefont{E.~D.} \bibnamefont{Bauer}}, \bibnamefont{and}
  \bibinfo{author}{\bibfnamefont{A.}~\bibnamefont{Kapitulnik}},
  \bibinfo{journal}{Phys. Rev. B} \textbf{\bibinfo{volume}{91}},
  \bibinfo{pages}{140506} (\bibinfo{year}{2015}), ISSN
  \bibinfo{issn}{1098-0121}, \eprint{1410.1479},
  \urlprefix\url{https://link.aps.org/doi/10.1103/PhysRevB.91.140506}.

\bibitem[{\citenamefont{Mekonen et~al.}(2022)\citenamefont{Mekonen, Kang,
  Chaudhuri, Barbalas, Ran, Kotliar, Butch, and Armitage}}]{Mekonen2022}
\bibinfo{author}{\bibfnamefont{S.~M.} \bibnamefont{Mekonen}},
  \bibinfo{author}{\bibfnamefont{C.-J.} \bibnamefont{Kang}},
  \bibinfo{author}{\bibfnamefont{D.}~\bibnamefont{Chaudhuri}},
  \bibinfo{author}{\bibfnamefont{D.}~\bibnamefont{Barbalas}},
  \bibinfo{author}{\bibfnamefont{S.}~\bibnamefont{Ran}},
  \bibinfo{author}{\bibfnamefont{G.}~\bibnamefont{Kotliar}},
  \bibinfo{author}{\bibfnamefont{N.~P.} \bibnamefont{Butch}}, \bibnamefont{and}
  \bibinfo{author}{\bibfnamefont{N.~P.} \bibnamefont{Armitage}},
  \bibinfo{journal}{Physical Review B} \textbf{\bibinfo{volume}{106}},
  \bibinfo{pages}{085125} (\bibinfo{year}{2022}), ISSN
  \bibinfo{issn}{2469-9950}, \eprint{2105.05121},
  \urlprefix\url{https://link.aps.org/doi/10.1103/PhysRevB.106.085125}.

\bibitem[{\citenamefont{Iguchi et~al.}(2022)\citenamefont{Iguchi, Man, Thomas,
  Ronning, Rosa, and Moler}}]{Iguchi2022}
\bibinfo{author}{\bibfnamefont{Y.}~\bibnamefont{Iguchi}},
  \bibinfo{author}{\bibfnamefont{H.}~\bibnamefont{Man}},
  \bibinfo{author}{\bibfnamefont{S.~M.} \bibnamefont{Thomas}},
  \bibinfo{author}{\bibfnamefont{F.}~\bibnamefont{Ronning}},
  \bibinfo{author}{\bibfnamefont{P.~F.~S.} \bibnamefont{Rosa}},
  \bibnamefont{and} \bibinfo{author}{\bibfnamefont{K.~A.} \bibnamefont{Moler}},
  pp. \bibinfo{pages}{1--12} (\bibinfo{year}{2022}), \eprint{2210.09562},
  \urlprefix\url{http://arxiv.org/abs/2210.09562}.

\bibitem[{\citenamefont{Opletal et~al.}(2022)\citenamefont{Opletal, Sakai,
  Haga, Tokiwa, Yamamoto, Kambe, and Tokunaga}}]{Opletal2022}
\bibinfo{author}{\bibfnamefont{P.}~\bibnamefont{Opletal}},
  \bibinfo{author}{\bibfnamefont{H.}~\bibnamefont{Sakai}},
  \bibinfo{author}{\bibfnamefont{Y.}~\bibnamefont{Haga}},
  \bibinfo{author}{\bibfnamefont{Y.}~\bibnamefont{Tokiwa}},
  \bibinfo{author}{\bibfnamefont{E.}~\bibnamefont{Yamamoto}},
  \bibinfo{author}{\bibfnamefont{S.}~\bibnamefont{Kambe}}, \bibnamefont{and}
  \bibinfo{author}{\bibfnamefont{Y.}~\bibnamefont{Tokunaga}}, pp.
  \bibinfo{pages}{1--7} (\bibinfo{year}{2022}), \eprint{2211.16760},
  \urlprefix\url{http://arxiv.org/abs/2211.16760}.

\bibitem[{\citenamefont{Tougait et~al.}(2001)\citenamefont{Tougait,
  Andr{\'{e}}, Bour{\'{e}}e, and No{\"{e}}l}}]{Tougait2001}
\bibinfo{author}{\bibfnamefont{O.}~\bibnamefont{Tougait}},
  \bibinfo{author}{\bibfnamefont{G.}~\bibnamefont{Andr{\'{e}}}},
  \bibinfo{author}{\bibfnamefont{F.}~\bibnamefont{Bour{\'{e}}e}},
  \bibnamefont{and}
  \bibinfo{author}{\bibfnamefont{H.}~\bibnamefont{No{\"{e}}l}},
  \bibinfo{journal}{Journal of Alloys and Compounds}
  \textbf{\bibinfo{volume}{317-318}}, \bibinfo{pages}{227}
  (\bibinfo{year}{2001}), ISSN \bibinfo{issn}{09258388},
  \urlprefix\url{https://linkinghub.elsevier.com/retrieve/pii/S0925838800013335}.

\bibitem[{\citenamefont{Fried et~al.}(2014)\citenamefont{Fried, Fejer, and
  Kapitulnik}}]{Fried2014b}
\bibinfo{author}{\bibfnamefont{A.}~\bibnamefont{Fried}},
  \bibinfo{author}{\bibfnamefont{M.}~\bibnamefont{Fejer}}, \bibnamefont{and}
  \bibinfo{author}{\bibfnamefont{A.}~\bibnamefont{Kapitulnik}},
  \bibinfo{journal}{Review of Scientific Instruments}
  \textbf{\bibinfo{volume}{85}}, \bibinfo{pages}{103707}
  (\bibinfo{year}{2014}), ISSN \bibinfo{issn}{0034-6748},
  \urlprefix\url{http://dx.doi.org/10.1063/1.4897184
  http://aip.scitation.org/doi/10.1063/1.4897184}.

\end{thebibliography}

\end{document}



\title{Supplemental Information for ``The fate of time-reversal symmetry breaking in UTe$_2$"}

\author{M.O. Ajeesh}
\author{M. Bordelon}
\author{C. Girod}
\author{S. Mishra}
\author{F. Ronning}
\author{E.D. Bauer}
\author{B. Maiorov}
\author{J.D. Thompson}
\author{P.F.S. Rosa}
\author{S.M. Thomas}
\email{smthomas@lanl.gov}
\affiliation{MPA-Q, Los Alamos National Lab, Los Alamos, NM 87544}

\date{\today}

\maketitle

\subsection{Additional Kerr data}
\begin{figure}[H]
	\begin{center}
		\includegraphics[width=1\columnwidth]{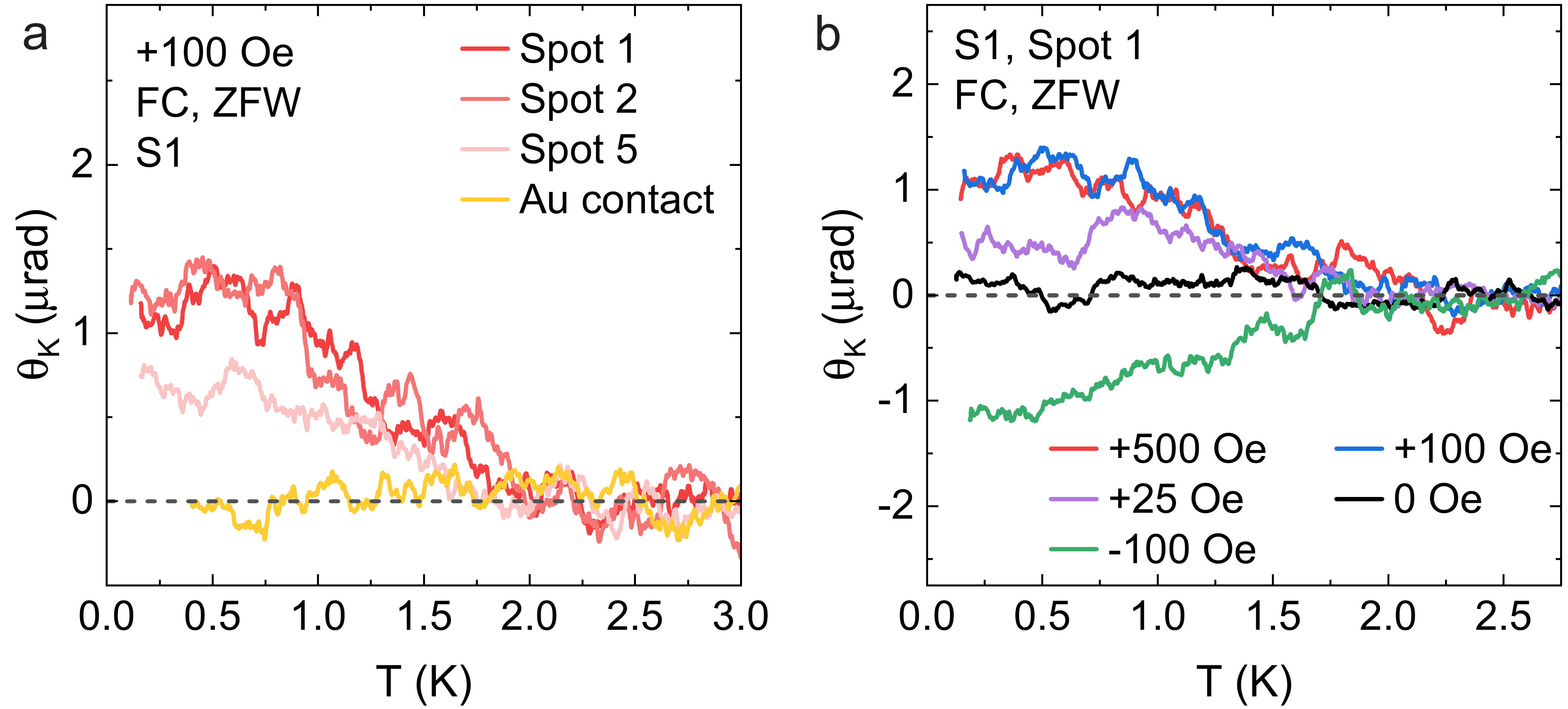}
		\caption{
			Additional data for sample S1.
			(a)~Zero-field-warmed $\theta_K$ after cooling in $+100$~Oe for different spots on sample S1. The gold contact shows a zero $\theta_K$ at all temperatures.
			(b)~Zero-field-warmed $\theta_K$ after cooling in variable fields for spot 1 on sample S1.
		}
		 \label{fig:sfig1}
	\end{center}
\end{figure}

\begin{figure}[H]
	\begin{center}
		\includegraphics[width=1\columnwidth]{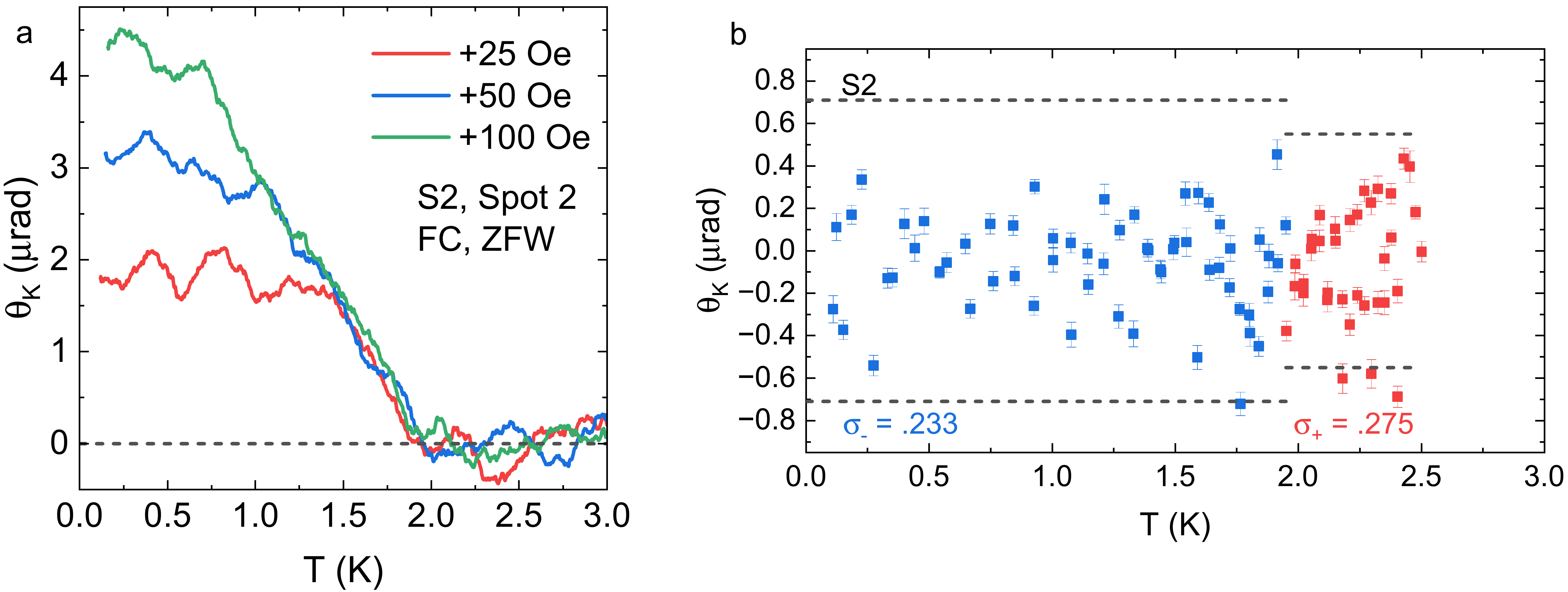}
		\caption{
			Additional data for sample S2:
			(a)~Zero-field-warmed $\theta_K$ after cooling in variable fields for spot 2 on sample S2.
			(b)~Measured value of $\theta{}_K$ as a function of temperature for sample S2. The sample is heated above T$_c$ between each data point. Red indicates T$>$T$_c$; blue indicates T$<$T$_c$. As sample resistivity was not available, the T$_c$ was inferred from temperature-dependent Kerr measurements. A similar amount of laser pulse power was used as in other measurements to ensure that the sample was heated above T$_c$ between each data point.
		}
		 \label{fig:sfig2}
	\end{center}
\end{figure}

\begin{figure}[H]
	\begin{center}
		\includegraphics[width=1\columnwidth]{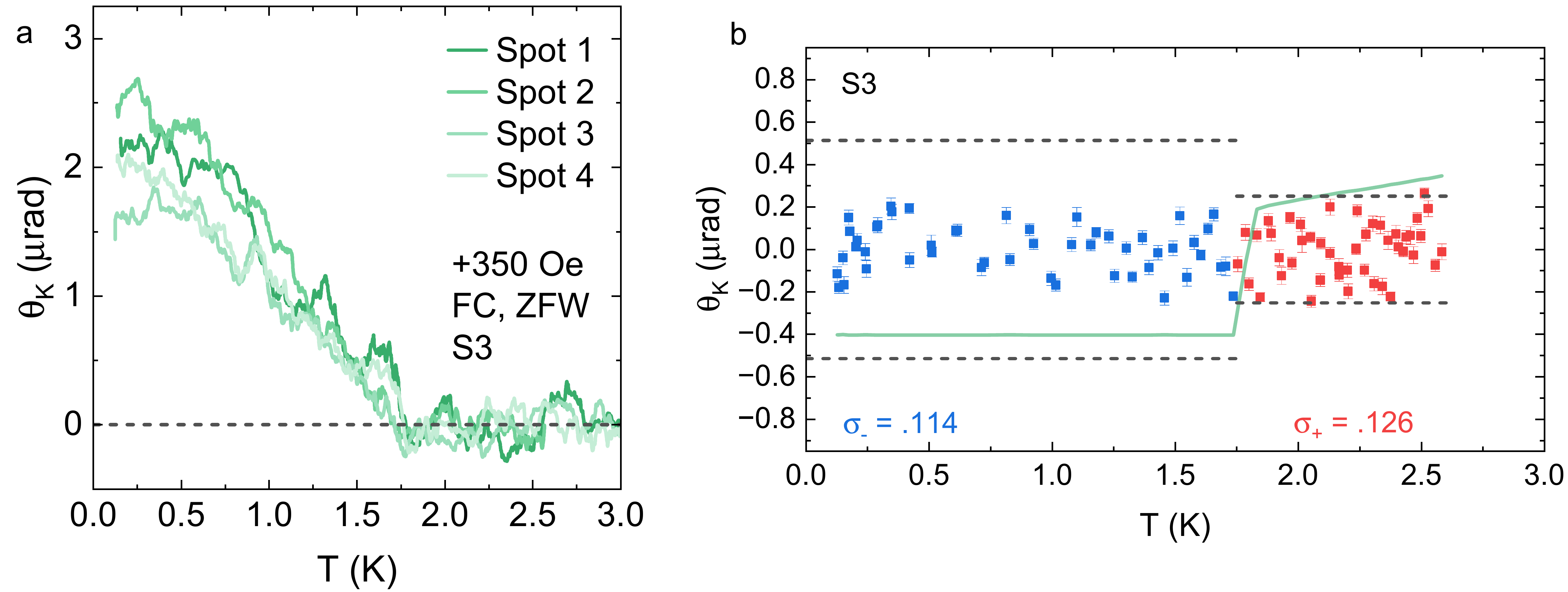}
		\caption{
			Additional data for sample S3:
			(a)~Zero-field-warmed $\theta_K$ after cooling in $+350$~Oe for different spots on sample  S3.
			(b)~Measured value of $\theta{}_K$ as a function of temperature for sample S3. The sample is heated above T$_c$ between each data point. Red indicates T$>$T$_c$; blue indicates T$<$T$_c$. The faint green line shows the sample resistivity.
		}
		 \label{fig:sfig3}
	\end{center}
\end{figure}

\begin{figure}[H]
	\begin{center}
		\includegraphics[width=1\columnwidth]{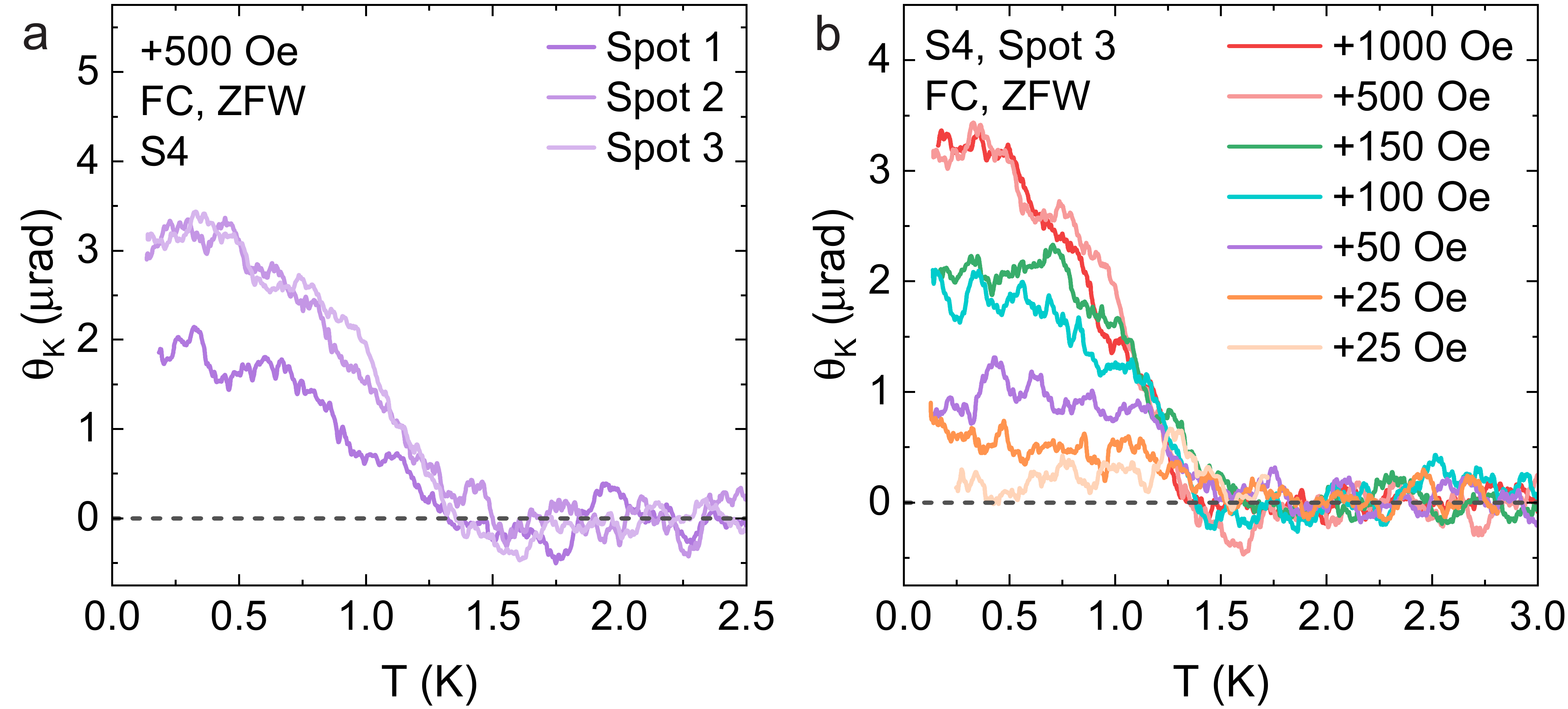}
		\caption{
			Additional data for sample S4.
			(a)~Zero-field-warmed $\theta_K$ after cooling in $+500$~Oe for different spots on sample S4.
			(b)~Zero-field-warmed $\theta_K$ after cooling in variable fields for spot 3 on sample S4.
		}
		 \label{fig:sfig3}
	\end{center}
\end{figure}

\begin{figure}[H]
	\begin{center}
		\includegraphics[width=1\columnwidth]{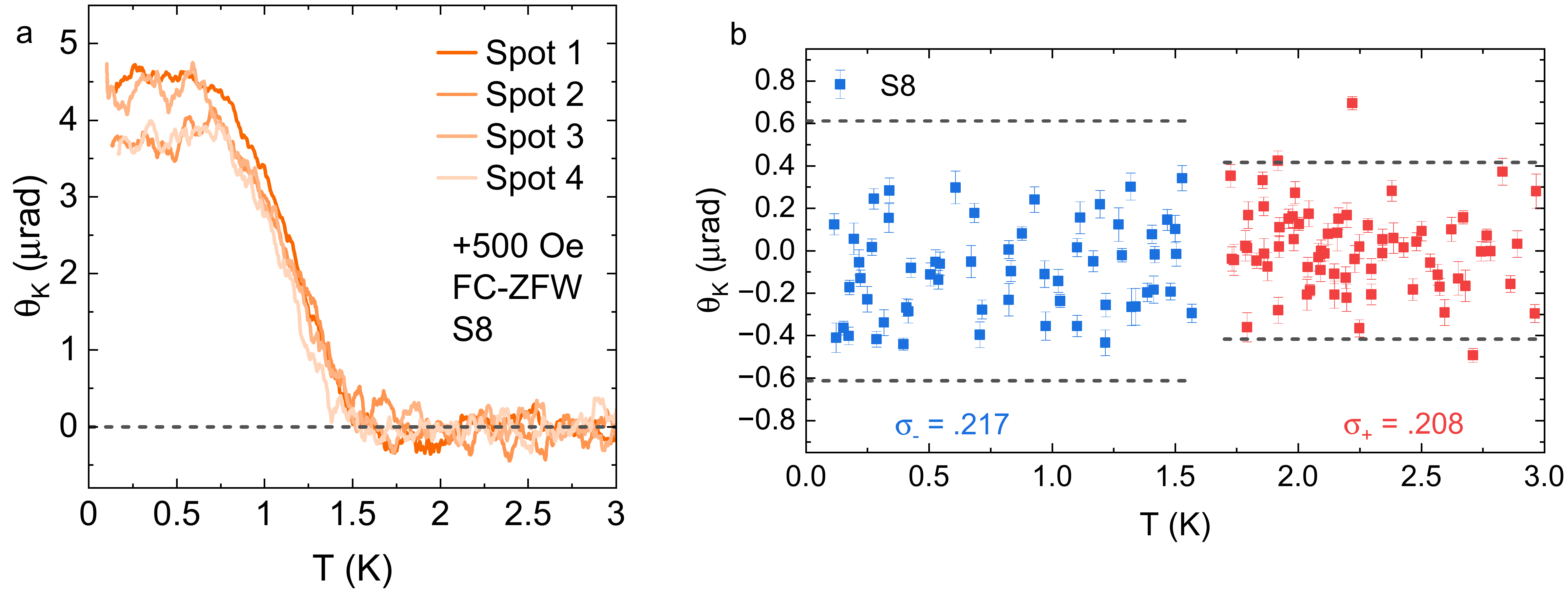}
		\caption{
			Data for sample S8.
			This sample was measured on an as-grown surface instead of a polished surface.
			(a)~Zero-field-warmed $\theta_K$ after cooling in $+500$~Oe for different spots on sample S8.
			(b)~Measured value of $\theta{}_K$ as a function of temperature for sample S8. The sample is heated above T$_c$ between each data point. Red indicates T$>$T$_c$; blue indicates T$<$T$_c$. The sample was heated using a resistive heater instead of a laser, which allowed for the sample to be cooled through T$_c$ more slowly over 60~seconds.
		}
		 \label{fig:sfig3}
	\end{center}
\end{figure}

\subsection{Normalization of ac susceptibility data}
\begin{figure}[H]
	\begin{center}
		\includegraphics[width=0.5\columnwidth]{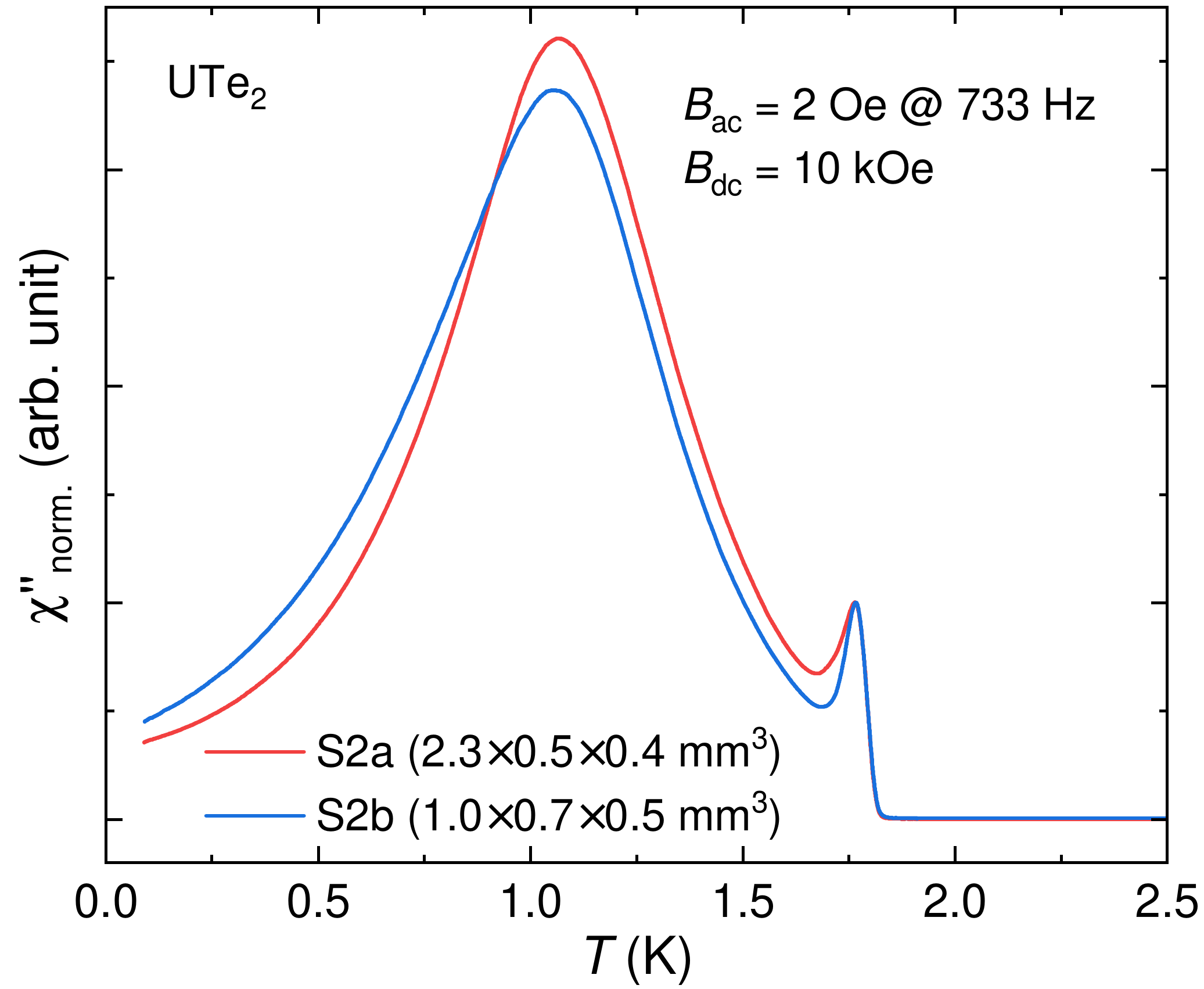}
		\caption{
			Normalized out-of-phase component of the ac susceptibility $\chi''(T)$ of two different pieces of sample S2.
			Data are normalized with respect to the height of the peak at the onset of superconductivity.
			B$_{dc}$ and B$_{ac}$ are applied along the length ($a$ axis) of the bar shaped samples.
		}
		 \label{fig:sfig3}
	\end{center}
\end{figure}